\documentclass[12pt,preprint]{aastex}

\begin{document}
\title{A Survey of {\it FUSE} and {\it HST} Sightlines through High-Velocity Cloud Complex C}
\author{Joseph A. Collins, J. Michael Shull\altaffilmark{1}}
\affil{University of Colorado, CASA, Department of Astrophysical \& Planetary Sciences, Campus Box 389, Boulder, CO 80309}
\altaffiltext{1}{Also at JILA, University of Colorado and National Institute of Standards and Technology.}
\and
\author{Mark L. Giroux}
\affil{East Tennessee State University, Department of Physics \& Astronomy, Box 70652, Johnson City, TN 37614}

\begin{abstract}

Using archival Far Ultraviolet Spectroscopic Explorer ({\it FUSE}) and 
Hubble Space Telescope ({\it HST}) data, we have assembled a survey of eight
sightlines through high-velocity cloud Complex C.  Abundances of the
observed ion species vary significantly for these sightlines, indicating that
Complex C is not well characterized by a single metallicity. Reliable 
metallicities based on [\ion{O}{1}/\ion{H}{1}] range from 0.1-0.25 solar.   
Metallicities based on [\ion{S}{2}/\ion{H}{1}] 
range from 0.1-0.6 solar, but 
the trend of decreasing abundance with 
\ion{H}{1}\ column density indicates that photoionization 
corrections may affect the conversion to [S/H].  We present models 
of the dependence of the ionization correction on \ion{H}{1}\ column
density;  these ionization corrections are significant
when converting ion abundances to elemental abundances for S, Si, and Fe. 
The measured abundances in this survey indicate that 
parts of the
cloud have a higher metallicity than previously thought and 
that Complex C may represent a mixture of ``Galactic fountain'' gas
with infalling low-metallicity gas.  We find that [S/O] and
[Si/O] have a solar ratio, suggesting little dust depletion.
Further, the measured abundances suggest
an over-abundance of O, S, and Si relative to N and Fe.  The enhancement of 
these $\alpha$-elements suggests that the bulk of the metals in Complex C were 
produced by Type II supernovae and then removed from the star-forming region,
possibly via supernovae-driven winds or tidal stripping, 
before the ISM could be enriched by N and Fe.

\end{abstract}

\section{Introduction}

Almost four decades after their discovery (Muller, Oort, \&
Raimond 1963), the origin of high-velocity clouds
(HVCs; Wakker \& van Woerden 1997) remains a mystery.
Defined by their incompatibility with a simple model of 
differential Galactic rotation, HVCs exhibit sub-solar metallicities and 
are currently favored to reside in the Galactic
halo.  Within this picture,
it is not clear whether the source of HVCs is predominantly the accretion of
low-metallicity gas onto the Galaxy (Wakker et al. 1999)
or whether the material traces condensed 
outflows of hot enriched gas in a ``Galactic fountain'' 
(Shapiro \& Field 1976; Bregman 1980). Intermediate-velocity clouds (IVCs),
in contrast, exhibit solar metallicities and 
are thought to be of Galactic origin (Richter et al. 2001).   
Alternative hypotheses for the HVCs 
identify the majority of them as extragalactic objects
within the Local Group at Galactocentric distances up to 500 kpc
(Blitz et al. 1999), or posit that at least the subset known
as Compact High Velocity Clouds (CHVCs) are extragalactic
(Braun \& Burton 1999).  These objects would then be remnants of the formation
of the Local Group.
The statistics of moderate-redshift
Mg II and Lyman limit absorbers (Charlton, Churchill, \& Rigby 2000),
would seem to rule out this hypothesis as a general scenario
for group formation.
In any event, the simulations which underpin the arguments for
an extragalactic origin for the HVCs generally do not require
such an identification for the very large complexes of HVCs,
such as Complex C. However, within
their model it is possible that Complex C might represent
one of these building blocks viewed at a very close distance.

Complex C occupies nearly the same angular extent ($\sim2000$ deg$^{2}$)
as the Magellanic 
Stream and consists of a number of separate HVC cores in the 
northern Galactic hemisphere.  The Magellanic Stream has been linked to the 
Small Magellanic Cloud (SMC) through
modeling of a tidal interaction with the Milky Way (Gardiner \& Noguchi 1996)
and through the similarity in metallicity of the Stream 
(0.2-0.3 solar; Lu et al. 1998; Gibson et al. 2000; Sembach et al. 2001) 
and SMC.  In contrast, Complex C shows no 
obvious connection to the Galaxy.  The distance to Complex C is at least 
1.5 kpc (de Boer et al. 1994) from the non-detection of absorption lines
toward high-latitude stars, although Wakker (2001) has recently 
put a new, albeit weak, lower limit at 6 kpc.  

The lack of a strong distance constraint or evidence for 
interaction with the Galaxy make absorption-line studies of abundances
in quasar sightlines the best option for gaining insight into the
origin of Complex C.  The first such study for Complex C, from {\it HST}  
Goddard High Resolution Spectrograph (GHRS) data along 
the Markarian 290 sightline by Wakker et al. (1999), inferred a metallicity of
0.1 solar based on the H$\alpha$-corrected [\ion{S}{2}/\ion{H}{1}] abundance.  
Those authors interpreted this 
abundance as evidence for infalling low-metallicity gas from the intergalactic
medium (IGM) posited to explain the ``G-dwarf problem.''  
Recently, however, other studies have demonstrated that the 
metallicity of Complex C varies from sightline to sightline.  Gibson et al. 
(2001; hereafter G01) looked at all available archival GHRS and 
Space Telescope Imaging Spectrograph (STIS) 
data for five Complex C sightlines (including Mrk 290) and found that the 
metallicity inferred from the \ion{S}{2} abundance varies 
between 0.1-0.4 solar, implying that Complex C may  
be contaminated by enriched gas from the Galactic disk.
Other recent studies complicate the picture and further illustrate the 
point that Complex C may not be well described by a single metallicity. 
Richter et al. (2001a) combine {\it FUSE} and STIS data 
for the PG 1259+593 sightline and find an \ion{O}{1}\ abundance of
0.1 solar, while Murphy et al. (2000) find an abundance of \ion{Fe}{2}\ 
near 0.5 solar from {\it FUSE} observations of the Mrk 876 sightline.

The increasing availability of high-quality, far-ultraviolet
{\it FUSE} and {\it HST}/STIS spectra has allowed the study of multiple
transitions of the dominant ionization stages of N, O, Si, Fe, Ar, P,
and S.  In addition to estimates of the absolute abundances
of these elements, the relative abundances of these elements
can be inferred, sometimes more precisely than their absolute abundances.
This allows a preliminary study of the depletion of
refractory elements onto dust grains and 
the nucleosynthetic history of the gas.

Using {\it HST}-GHRS spectra, Savage \& Sembach (1996) find that gas-phase
abundances of refractory elements in the Galactic halo can be significantly
depleted.  For five halo sightlines, those authors find that the mean
depletion from solar abundance is $\sim$ 0.3 dex for Si and $\sim$ 0.6 dex for
Fe, while the S abundance is essentially solar.  
However, it is difficult to completely separate the effects of depletion
from ionization effects. Although each element has a dominant
ionization stage, no element is completely in one stage, and the
fraction remaining in other stages varies with the element.  As we
show below, with enough measurements of different elements, it
is possible to separate these two effects and assess the degree
of depletion along sightlines toward Complex C.

Recent reviews have discussed the long history of exploiting 
relative abundances to reveal information about the chemical
evolution of the gas (e.g., Pettini 2002; Matteucci 2002).
Alpha-process elements (such as  
O, Ne, Si, and S) are produced predominantly by Type II supernovae (SNe), 
while Fe is produced primarily by Type Ia SNe.  Nitrogen, however, is primarily
produced by lower mass stars in the asymptotic giant branch (AGB) stage.  With
the bulk of $\alpha$-element production occurring relatively soon after 
star-formation is initiated, the release of N and Fe into the ISM can thus 
lag S or O enrichment by $\sim250$ Myr 
(Henry et al. 2000).  If the gas is somehow stripped from star-forming 
regions, either by a tidal interaction or in an outflow induced 
by combined SN-activity, then the resulting metal-content will be
significantly N-poor.  G01 and Wakker (2001) each interpret the low
relative abundance of N in Complex C as evidence for primary enrichment by
Type II SNe.

The cited Complex C abundance studies have all used different methods to 
infer metallicities of Complex C.  Some studies fit the Complex C metal-line 
profiles
with a Gaussian to determine the equivalent width, $W_{\lambda}$, while others
simply integrate the profile over the Complex C velocity range established by 
the \ion{H}{1}\ profile.  These studies have also used different 
methods to determine species' column densities, such as curve-of-growth (CoG)
fitting, an apparent optical depth (AOD) calculation (Savage \&
Sembach 1991), or an assumption
that these profiles are optically thin.  In addition, except 
for the Richter et al. (2001a) work, these studies have typically concentrated 
on either {\it FUSE} or {\it HST} data alone.  
The {\it FUSE} data have the advantage that the covered 
spectral range includes numerous lines of \ion{O}{1}, whose abundance is
not as strongly influenced as \ion{S}{2}\ by ionization effects
(charge exchange keeps \ion{O}{1}\ closely coupled to \ion{H}{1}).

The goal of this work is to study all 
archival {\it FUSE} and {\it HST} data for sightlines through Complex C. Using
a consistent method of analysis, we then 
determine abundances for the various metal
species to infer information regarding the origin of HVC Complex C.  
The positions of the eight sightlines are shown in Figure 1, overlayed
on \ion{H}{1}\ emission data for Complex C from the Leiden-Dwingeloo Survey
(LDS; Hartmann \& Burton 1997). 
The {\it FUSE} and {\it HST} observations are discussed in \S\ 2.  
Results for the 
analysis for the eight sightlines are presented in \S\ 3.  The results are 
discussed in \S\ 4.

\section{Observations}

\subsection{The {\it FUSE} data}

The {\it FUSE} observations for these eight sightlines were obtained from the 
publicly available Multimission Archive (MAST) at the Space Telescope Science 
Institute. The total exposure times for {\it FUSE} observations towards the
sightlines range from 8.3 ks for PG 1626+554 to 610.6 ks for PG 1259+593 and, 
as a result, data quality within the sample varies significantly.  All 
sightlines were observed through the $30\arcsec \times30\arcsec$ LWRS 
aperture in time-tag mode. A summary of the {\it FUSE} 
observations is shown in 
Table 1.  For a complete description of the {\it FUSE} instrument and its 
operation see Moos et al. (2000) and Sahnow et al. (2000).

\placetable{t1}

The raw data were obtained from the archive and calibrated spectra were 
extracted using a pass through the CALFUSE 
(v. 2.0.5) reduction pipeline.  
This version of the pipeline makes the correct transformation to a 
heliocentric wavelength scale.  In versions earlier than 2.0, a sign error 
in file-header keywords resulted in the wrong heliocentric wavelength 
correction.  In order to improve signal-to-noise for these data we include 
both ``day'' and ``night'' photons in the final calibrated spectra.  The 
inclusion of ``day'' photons leads to strong airglow contamination of 
interstellar \ion{O}{1}\ and \ion{N}{1}\ absorption lines.  However, the 
contamination is generally centered at $V_{LSR}=0$ km s$^{-1}$ 
and does not affect the 
higher-velocity Complex C absorption as confirmed by comparisons of 
unscreened spectra to spectra where ``day'' 
photons were screened by CALFUSE.  
Individual exposures were then coadded, weighted by their exposure times, to 
yield a final spectrum.

A single resolution element of the {\it FUSE} spectrum is about 20 km s$^{-1}$
($\sim10$ pixels) and, as a result 
data are oversampled at that resolution.  Thus, to further improve 
signal-to-noise, data were rebinned over 3, 5, or 7 pixels depending on the 
initial data quality.  A common problem with analysis of {\it FUSE} spectra is 
the setting of an absolute wavelength scale.  The most common technique is 
to shift the wavelength scale by determining the velocity offset of 
line centroids from known absorption features.  This was carried out by 
comparing the centroid of Galactic \ion{H}{1}\ emission to various Galactic 
absorption features such as those of \ion{Ar}{1}\ ($\lambda$1048.22, 
1066.66), \ion{Fe}{2} ($\lambda$1125.45, 1144.94), and H$_{2}$.  We 
estimate the absolute wavelength scale to be accurate to within $\sim10$ 
km s$^{-1}$.

\subsection{The {\it HST} data}

All available calibrated {\it HST} data for these 
sightlines were also obtained 
from MAST.  The data consist of STIS 
and GHRS observations for seven of the 
sightlines.  No {\it HST} spectroscopic data exist 
for PG 1626+554.  See Table 2 
for a summary of the {\it HST} observations.  Final spectra were obtained by 
coadding individual exposures, weighted by their exposure times.

\placetable{t2}

In order to both improve signal-to-noise and to match pixel size in velocity 
to that of the rebinned {\it FUSE} data, 
the {\it HST} data were rebinned to 3 or 5 
pixels.  In a few cases, data were not rebinned as the pixel sizes were 
already comparable to the rebinned {\it FUSE} data.  Finally, the absolute 
wavelength scale was set as for the {\it FUSE} data, comparing the \ion{N}{1}\ 
($\lambda$1199.550) and \ion{S}{2}\ ($\lambda$1250.584, 1253.811, 1259.519) 
lines to the centroid of Galactic \ion{H}{1}\ emission.

\section{Results}

We detect metal-line absorption associated with Complex C in each of these 
sightlines.  Individual line profiles were normalized by fitting low-order 
polynomials to the continuum in the region immediately surrounding the line 
in question.  Regions within $\pm3$ \AA\ about the Galactic line-centroid were 
used for the continuum normalization, although in a number of cases spurious 
absorption near the line required the use of a much larger region for 
continuum measurement.  For each sightline, the velocity range of Complex C 
is determined from the \ion{H}{1}\ profile from the LDS data.  
The equivalent width, $W_{\lambda}$, of an 
absorption feature could then be measured by integrating the line over the 
velocity range occupied by Complex C.  In the cases of 
\ion{Fe}{3}\ $\lambda$1122.524 and \ion{N}{1}\ $\lambda$1134.165, the complex C
components are contaminated by the Galactic components of the \ion{Fe}{2}\ 
$\lambda$1121.975 and $\lambda$1133.665, respectively.  The contributions of 
these \ion{Fe}{2}\ lines to the absorption feature were determined from a 
CoG fit to other \ion{Fe}{2}\ lines integrated over the same 
velocity range.  
Uncertainties determined for the 
equivalent widths include error contributions from both photon statistics 
and continuum placement.  In some cases, photon counting errors were estimated 
empirically from the data.  For {\it FUSE} data, 
where individual lines are often 
measured in two different channels, equivalent widths are calculated by 
averaging values measured from the individual channels, weighted by a factor 
determined by their relative errors.  When lines can not be measured, we 
estimate 3$\sigma$ upper limits to the line equivalent width.  

In order to measure column densities of the various metal species, 
we fitted a
CoG to the data.  This requires the measurement 
of at least two lines for one species in each sightline.  We are able to 
empirically determine a best-fit CoG for five of the eight sightlines.  The 
three sightlines for which a CoG could not be fitted
to the data have the lowest 
signal-to-noise, and the lines for which $W_{\lambda}$ can
be measured are highly uncertain.  Thus, to estimate column densities, and 
more often than not their upper limits, we adopt a CoG with 
doppler parameter $b=10$ km s$^{-1}$ for the lower quality data.  This 
$b$-value is generally consistent with those obtained from the sightlines with
the highest quality data.  The 
doubly-ionized species of S and Fe typically reside in a different phase 
of the ISM than neutral and singly-ionized species and are thus not well 
described by a $b$-value determined from species of a lower degree 
of ionization. \ion{S}{3}\ and \ion{Fe}{3}\ in these sightlines have low 
measured values of $W_{\lambda}$, and their column densities are 
measured in the optically thin case (Spitzer 1978) by,
\begin{equation}
N(\mbox{cm}^{-2})=1.13\times10^{17}\frac{W_{\lambda}(\mbox{m\AA})}{f\lambda_{0}^{2}(\mbox{\AA})}.
\end{equation}
We adopt rest wavelengths and oscillator strengths from Morton (2002) except 
for several of the \ion{Fe}{2} lines, where the oscillator strengths are 
from Howk et al. (2000).

A significant problem in determining the abundances of these various ions
is in obtaining the \ion{H}{1}\ column density of Complex C 
along these sightlines.  
\ion{H}{1}\ columns cannot be determined from {\it HST} or {\it FUSE}
data, as neutral hydrogen absorption in the Lyman series from
Complex C is saturated and blended with saturated absorption from the Galaxy.  
A common approach in abundance studies is to adopt \ion{H}{1}\ columns from 
21-cm emission.  However, beam sizes in the single-dish data are, at the 
smallest, of order a few arc-minutes compared to the subarc-second effective 
beams in {\it HST} and {\it FUSE} quasar absorption-line studies.   
Even over arc-minute 
scales, variations in \ion{H}{1} columns can be significant (Wakker \& 
Schwarz 1991) and if \ion{H}{1}\ is clumped on scales of arc-seconds, then 
single-dish data will poorly reflect columns along quasar sightlines.   
Our approach then, is to adopt \ion{H}{1}\ columns from the data of Wakker 
et al. (2001) taken at the Effelsberg 100-m telescope which, with a 
$9\arcmin.1$ beam, samples \ion{H}{1}\ columns on a far smaller scale than 
the LDS 36$\arcmin$ beam.  That survey includes \ion{H}{1}\ 
profiles for all of our Complex C 
sightlines except PG 1626+554, where we measure the 
\ion{H}{1}\ column from the LDS data.  For these sightlines, the \ion{H}{1}\ 
columns from the $9\arcmin.1$ beam range from 0.19 dex below to 0.28
dex above those of the 36$\arcmin$ beam.  We adopt this range as the
systematic error associated with the beam-size mismatch, 
although it should be stressed that the 
difference could be even greater when extrapolating to smaller scales.  
Future \ion{H}{1}\ observations of Complex C at high spatial-resolution would
be extremely useful to resolve this issue. 

Metallicities are measured with respect to the solar (meteoritic) abundances 
of Grevesse \& Sauval (1998; hereafter GS98)\footnotemark 
\footnotetext{The adopted solar
abundances from GS98 are as follows: 
[N/H]$=(8.3\pm1.2)\times10^{-5}$; [O/H]$=(6.8\pm0.9)\times10^{-4}$;
[Al/H]$=(3.1\pm0.1)\times10^{-6}$; [Si/H]$=(3.6\pm0.1)\times10^{-5}$;
[P/H]$=(3.6\pm0.5)\times10^{-7}$; [S/H]$=(1.6\pm0.2)\times10^{-5}$; 
[Ar/H]$=(2.5\pm0.4)\times10^{-6}$; [Fe/H]$=(3.2\pm0.1)\times10^{-5}$.
}.  In addition, measured abundances from previous metallicity
studies of Complex C
are revised using the GS98 abundances so as to make reliable 
comparisons to our results.   
Holweger (2001) has recently presented new solar photospheric
abundances, the most notable of which is a low value of
[O/H]$_{\sun}=5.45\times10^{-4}$.  The use of this abundance\footnotemark 
\footnotetext{Throughout 
this work, for a species X this quantity is defined as, 
[$X$/H I]$_{\rm C}$=log$\left(\frac{N(X)/N(HI)_{\rm C}}{A(X)_{\sun}}\right)$, 
where 
$A(X)_{\sun}$ is the solar abundance of element X.  The subscript ``C''
refers to the measurement for Complex C.} 
would increase our measured values of [\ion{O}{1}/\ion{H}{1}]$_{\rm C}$ 
by 0.094 dex. We use \ion{O}{1}\ as a probe of cloud 
metallicity because 
the ionization state of oxygen is coupled to hydrogen through 
a resonant charge exchange reaction.  Previous studies (Wakker et al. 1999; 
G01) have used [\ion{S}{2}/\ion{H}{1}]$_{\rm C}$ to 
probe metallicity, but have had to contend with ionization corrections, since,
as we discuss in \S\ 4.1, that quantity seems to 
depend on the photoionization correction
as well as the actual sulfur metallicity.

In a number of cases we are unable to determine a metallicity based on
[\ion{O}{1}/\ion{H}{1}]$_{\rm C}$ due to contamination of the \ion{O}{1}\
$\lambda$1039.23 line by Galactic H$_{2}$ (5-0) R(2) Lyman absorption.  In 
addition, the data quality for these particular sightlines precludes 
measurements of the weaker \ion{O}{1}\ lines in the 920-950 \AA\ range of
the {\it FUSE} SiC channels.  We can estimate the column density of 
\ion{O}{1}\ along a given sightline at 
$N_{15}$(\ion{O}{1})$=(0.68)Z_{0.1}N_{19}$(\ion{H}{1}), where $N_{n}$ is the 
column density in units of $10^{n}$ cm$^{-2}$ and $Z_{0.1}$ is the metallicity 
in units of 0.1 solar.  Thus in cases where [\ion{O}{1}/\ion{H}{1}]$_{\rm C}$
could not be measured due to H$_{2}$ contamination, 
we estimate the expected equivalent width of the
\ion{O}{1}\ $\lambda$948.686 line based on the derived or assumed CoG for each
sightline.

We have also made an attempt to measure molecular hydrogen in Complex C.  
Richter et al. (2001b, 2002) previously 
analyzed {\it FUSE} data for the PG 1259+593 and PG 1351+640 sightlines
and were unable to detect high velocity H$_{2}$ associated with Complex C, 
whereas they find that molecular hydrogen
is ubiquitous in IVCs where dust is more prevalent.  
In many cases we detect strong Galactic H$_{2}$ absorption, although H$_{2}$
associated with Complex C could not be measured within the 
sensitivity limits of the observations. {\it FUSE} is typically sensitive
to H$_{2}$ column densities as low as $\sim10^{14}$ cm$^{-2}$.
In order to put upper limits on 
column densities of H$_{2}$, we measure 3$\sigma$ limits on $W_{\lambda}$ 
for the strong $J=0$ and $J=1$ Werner lines and assume that the lines 
are optically thin.  Upper limits on molecular 
hydrogen column density as well as the molecular fraction, 
$f$(H$_{2}$)=2$N$(H$_{2}$)/[2$N$(H$_{2}$)+$N$(\ion{H}{1})], are shown in 
Table 3.  Clearly there is very little molecular gas associated with Complex C.

\placetable{t3}

Spectra for these eight sightlines, as well as line equivalent widths and 
the resulting column densities, abundances, and metallicities of the 
observed species are presented in \S\S\ 3.1-3.8.  Line profiles for \ion{O}{6}\
$\lambda$1031.93 are presented but not discussed.  For a thorough 
discussion of high-velocity \ion{O}{6}\ from these and other sightlines see 
Sembach et al. (2002).

\subsection{Mrk 279}

Mrk 279 has good quality {\it FUSE} 
data with an average S/N of $\sim9$ per pixel 
in the 1040-1050 \AA\ range in the LiF1a channel.  The GHRS data have been 
previously analyzed by G01, who determined a Complex C sulfur 
metallicity, revised assuming a GS98 solar abundance,
 of [\ion{S}{2}/\ion{H}{1}]$_{\rm C}=-0.29\pm0.18$ based 
on the measurement of \ion{S}{2}\ $\lambda$1250 alone.

Profiles of some of the observed species are shown in Figure 2, along with 
the Effelsberg profile of \ion{H}{1}\ emission.  
High-velocity \ion{H}{1}\ emission 
is clearly present, though there appears to be significant blending with 
intermediate 
velocity components.  Wakker et al. (2001) identify eight components to 
the full \ion{H}{1}\ profile, two of which are at high velocity.  We  
follow the approach of G01 and choose an integration range 
that spans these two components ($-180\leq V_{LSR}\leq-90$ km s$^{-1}$).
Since there
appears to be significant blending with an IVC, we directly integrate the 
Effelsberg \ion{H}{1}\ profile to determine $N$(\ion{H}{1}) over this range.

Equivalent widths for the Mrk 279 sightline are shown in Table 4.  The
\ion{S}{2} $\lambda$1253 line could not be measured from the GHRS spectrum 
owing to contamination from a Ly$\alpha$ absorber intrinsic to Mrk 279.  The
\ion{S}{2}\ $\lambda$1259 line is not included in the spectrum, and thus only
\ion{S}{2} $\lambda$1250 is reported.  G01 raise the 
issue that an intrinsic Ly$\alpha$ absorber seen just redward of \ion{S}{2}\ 
$\lambda$1250 may contaminate the line at higher negative velocities, although
the equivalent width of the Complex C component and the implied \ion{S}{2}\
metallicity are consistent with other sightlines in our survey.  

\placetable{t4}

We have measured the equivalent widths of five \ion{O}{1}\ and four 
\ion{Fe}{2}\
lines associated with Complex C.  The data are best fitted 
by a CoG with doppler 
parameter $b=9.7^{+5.0}_{-2.8}$ km s$^{-1}$, as shown in Figure 3.  Resulting 
column densities, abundances, and metallicities for the ion species are
shown in Table 4.  The implied metallicity for this sightline 
from the \ion{O}{1}\ measurement is around $\sim0.2$ solar, significantly 
lower than the $\sim0.5$ solar measurement of G01 from 
\ion{S}{2}\ data.  We measure a \ion{S}{2}\ metallicity closer to $\sim0.6$ 
solar from the same data, and we suspect that 
ionization effects may play a role
for that species (see \S\ 4.1). 

We note that the \ion{H}{1}\ profile seems to indicate a significant blending
of the lower-velocity Complex C component (centered at $V_{LSR}\sim-105$ km 
s$^{-1}$) with an IVC (centered at $V_{LSR}\sim-75$ 
km s$^{-1}$) known as IV9.  IVCs typically have metallicities near solar, and 
our chosen integration range for Complex C may include absorption from IV9,
thus biasing our measurements to higher metallicities. 
In order to test this, we have performed an abundance analysis of the 
higher-velocity Complex C component (centered at $V_{LSR}\sim-140$ km 
s$^{-1}$) using the integration range 
$-180\leq V_{LSR}\leq-120$ km s$^{-1}$.  Results from this analysis are shown 
in Table 4, where values in parenthesis indicate measurements for the
$-140$ km s$^{-1}$ component.  The abundances of this component are  
similar to that of both components in the full integration range.  The 
indicated metallicity based on \ion{O}{1}\ is $\sim15$\% lower than in the case
of the full integration range, indicating some contamination by IV9 in the
range $-120\leq V_{LSR}\leq-90$ km s$^{-1}$.  However, it should be noted that
the \ion{O}{1}\ abundance of this component indicates a metallicity of 
$\sim0.17$ solar.

\subsection{Mrk 290}

The GHRS data for this sightline have been 
well studied by Wakker et al. (1999) and 
G01.  Each group uses \ion{S}{2}\ measurements to infer
Complex C metallicities of about $\sim0.1$ solar.  
The {\it FUSE} data are poor,
consisting of only 13 ks of exposure time, and have an average S/N of only
$\sim1$ per unbinned pixel in the 1040-1050 \AA\ range in the LiF1a channel.

Profiles of some of the observed species are shown in Figure 4, along with 
the LDS profile of \ion{H}{1}\ emission.  The LDS \ion{H}{1}\ profile
shows a clear demarkation between the Complex C and Galactic \ion{H}{1}\
components.  The integration range for equivalent width determinations is thus
set by the extent of the Complex C component ($-165\leq V_{LSR}\leq-80$ km 
s$^{-1}$). The column density of \ion{H}{1}\ along this sightline from the
Effelsberg data is ($12.3\pm0.6$)$\times$10$^{19}$ cm$^{-2}$, the highest 
column sightline from this sample.  Wakker et al. (1999) combined Effelsberg
single-dish data with Westerbork Synthesis Radio Telescope (WSRT) 
interferometric data to measure $N$(\ion{H}{1})$=9.2\times10^{19}$ cm$^{-2}$ at
1$\arcmin$ resolution.  The high \ion{H}{1}\ column, 
coupled with the poor quality 
of the currently available {\it FUSE} data, 
makes this sightline an obvious choice
for further {\it FUSE} observations.

Given the data quality, we are only able to place upper limits on Complex
C metallicities for species in the {\it FUSE} bandpass.  In addition, strong 
Galactic H$_{2}$ (5-0) R(2) Lyman 
absorption blends with \ion{O}{1}\ absorption from Complex C, 
limiting our measurement of that species to only an upper limit.  
Equivalent width measurements, column densities, abundances, and 
metallicities for the sightline are summarized in Table 5.

\placetable{t5}

In order to measure column densities, we assume that the data can be fitted 
by a $b=10$ km s$^{-1}$ CoG, which adequately fits the \ion{S}{2}\ 
GHRS measurements.  Since the upper limits on the {\it FUSE} 
lines are quite high,
located on the flat part of the CoG, we use equation 1 to calculate column
densities from lines which are generally optically thin for the other 
sightlines, namely \ion{N}{1}\ $\lambda$1134.17, \ion{Si}{2}\ 
$\lambda$1020.70, and \ion{Ar}{1}\ $\lambda$1048.22. The \ion{O}{1}\
$\lambda$1039.23 and \ion{Fe}{2}\ $\lambda$1144.94 lines do not satisfy the 
optically thin requirement for the other sightlines, and 
column densities from these lines are derived via the CoG method.  

We obtain a similar \ion{S}{2}\ metallicity as Wakker et al. (1999) and G01 
for Complex C along this sightline, 
[\ion{S}{2}/\ion{H}{1}]$_{\rm C}=$$-1.00^{+0.16}_{-0.15}$.  
Using the WSRT \ion{H}{1}\ column would increase this value by 0.13 dex.
Again, it is likely that
the \ion{S}{2}\ metallicity is not the ideal tracer of the true metallicity
along this sightline.  Measurements of the H$_{2}$/\ion{O}{1}\ 
$\lambda$1039.23 blend place an unrestrictive 
upper limit to the metallicity along this sightline of $<0.6$ solar.
It would be useful to study \ion{O}{1}\ lines in the $920-950$ \AA\ range
to better constrain the metallicity.   
However, {\it FUSE}
observations of this wavelength range require 
long exposures owing to 
the relatively low-sensitivity of the SiC1b and SiC2a channels.
If the metallicity along this sightline lies in the range 0.1-0.25 solar,
then the expected $W_{\lambda}$ of the \ion{O}{1}\ $\lambda948.69$ line,
determined from the CoG, is within the range 100-115 m\AA.

\subsection{Mrk 501}

Existing {\it FUSE} data for this sightline are of 
generally poor quality with a S/N
of $\sim1.5$ per unbinned pixel in the $1040-1050$ \AA\ 
range of the LiF1a channel.  Due to the low signal-to-noise
of the {\it HST} data, G01 were able to place 
only upper limits on \ion{S}{2}\ metallicity.

Figure 5 shows some of the observed absorption profiles, along with the LDS
profile of \ion{H}{1}\ emission.  The high-velocity component of the 
\ion{H}{1}\ profile is not well-separated from Galactic and intermediate
\ion{H}{1}\ emission.  The only sensible approach is to integrate over
the two high-velocity \ion{H}{1}\ components 
($-135\leq V_{LSR}\leq-65$ km s$^{-1}$) identified by Wakker et al. (2001).  
The \ion{H}{1}\ column density, ($1.6\pm0.1$)$\times$10$^{19}$ cm$^{-2}$,
of Complex C along this sightline is the 
lowest among the sightlines in this sample.  

Since Mrk 501 absorption features in both the {\it FUSE} and 
GHRS spectra cannot be 
measured above the noise, we place 3$\sigma$ upper limits on
$W_{\lambda}$ for Complex C absorption features.  Equivalent width, 
column density, abundance, and 
metallicity upper limits for the sightline are summarized in Table 6.
The determination of column densities was carried out through a 
procedure analogous to that of the Mrk 290 data.

\placetable{t6}

The upper limit of [\ion{O}{1}/\ion{H}{1}]$_{\rm C}<$$-0.54$ places a 
useful 
limit on the metallicity at $<0.3$ solar.  With slightly 
improved signal-to-noise through a longer {\it FUSE} exposure 
for this sightline, the \ion{O}{1}\ metallicity could
possibly be measured.  Toward this end, we have planned a 55 ks exposure
with {\it FUSE} during Cycle 3.  
The limit we place on the \ion{S}{2}\ metallicity of
[\ion{S}{2}/\ion{H}{1}]$_{\rm C}<0.05$, near solar, is about 0.2 dex higher 
than the limit of G01.  

\subsection{Mrk 817}

The {\it FUSE} data for Mrk 817 are of high quality, 
with a S/N of $\sim11$ per 
unbinned pixel in the 1040-1050 \AA\ range for the LiF1a channel.  The GHRS
data are also of good quality and were analyzed by G01,
who measured a metallicity, revised assuming a GS98 solar abundance, 
of [\ion{S}{2}/\ion{H}{1}]$_{\rm C}=$$-0.41\pm0.07$ 
based on
measurements of the \ion{S}{2}\ $\lambda$1250 and $\lambda$1253 lines.

Figure 6 shows profiles of some of the observed species, as well as the LDS 
profile of \ion{H}{1}\ emission.  The Complex C component of the \ion{H}{1}\ 
profile is clearly separated from the Galactic and intermediate-velocity 
components, and the range of integration is chosen to match the velocity
range of the high-velocity component ($-140\leq V_{LSR}\leq-80$ km s$^{-1}$).

The measured values of $W_{\lambda}$ for the Mrk 817 sightline are shown in 
Table 7.  The \ion{S}{2}\ $\lambda$1250 and $\lambda$1253 lines 
are easily detected in the GHRS spectrum, although the stronger \ion{S}{2}\
$\lambda$1259 line is just redward of the spectrum's upper-wavelength cutoff.  
Determination of the local continuum associated with
the \ion{S}{2}\ $\lambda$1253 line is difficult owing to the superposition
of the line with the Ly$\alpha$ emission line intrinsic to Mrk 817.  The 
error associated with the uncertain continuum is not included in the
value quoted for \ion{S}{2}\ $\lambda$1253 in Table 7.

\placetable{t7}

We fitted a CoG with doppler parameter $b=10.8^{+3.2}_{-2.2}$ km s$^{-1}$,
shown in Figure 7, to equivalent 
width data for the five \ion{O}{1}, six \ion{Fe}{2}, and 
two \ion{S}{2} lines. 
The resulting column densities, abundances, and metallicities 
for the ion species are shown in Table 7.  The \ion{O}{1} measurement implies
a metallicity along this sightline of $\sim0.25$ solar, nearly the same
as that of the Mrk 279 sightline.  
Our measured value of [\ion{S}{2}/\ion{H}{1}]$_{\rm C}=$$-0.34\pm0.08$ 
implies a metallicity closer to $\sim0.5$ solar, although as for Mrk 279 a
photoionization correction may be necessary to obtain a true metallicity from
\ion{S}{2}\ measurements.  It should be noted that, although the 
\ion{O}{1}\ metallicity determined for the Mrk 817 sightline is consistent 
with the loose upper limit set for the Mrk 290 sightline, G01 
argue that their disparate \ion{S}{2}\ measurements cannot be reconciled 
by a single metallicity for Complex C.  

\subsection{Mrk 876}

Archival {\it FUSE} data for Mrk 876 
are of good quality, with a S/N of $\sim6$ per
unbinned pixel in the 1040-1050 \AA\ range of the LiF1a channel.  
Unfortunately, this is not an optimal sightline for metallicity studies as
it passes through a very large column of Galactic molecular hydrogen with
N(H$_{2}$)=2.3$\times$10$^{18}$ cm$^{-2}$ (Shull et al. 2000), 
strong absorption from which 
contaminates many of the most important Complex C absorption features.  
These data were originally presented by Murphy et al. (2000) who measured a 
metallicity, revised assuming a GS98 solar abundance, 
of [\ion{Fe}{2}/\ion{H}{1}]$_{\rm C}=$$-0.31^{+0.15}_{-0.23}$.  
The STIS
data for Mrk 876 include the \ion{N}{1} $\lambda1199.55$ line, although the 
\ion{S}{2}\ lines are not covered by the spectral bandpass.
G01 measured a surprisingly low \ion{N}{1}\ metallicity 
for this sightline and argue for a differing nucleosynthetic history
for the enrichment of sulfur and nitrogen.
 
Observed absorption line profiles are shown in Figure 8.  The Effelsberg  
profile of
\ion{H}{1}\ emission, also shown in Figure 8, exhibits two high-velocity 
components in the velocity range $-210\leq V_{LSR}\leq-95$ km s$^{-1}$.  This 
span is chosen as the integration range for the Complex C 
$W_{\lambda}$ measurements shown in Table 8. 
Owing to the 
strong Galactic H$_{2}$ absorption, we are unable to measure, or put
meaningful upper limits on, the
important \ion{O}{1}\ $\lambda$1039.23 line.  Other oxygen lines in the
920-950 \AA\ range are difficult to measure within the sensitivity limits, 
partly due to the absence of data for the SiC2 channel.  

\placetable{t8}

From the available
data, we are able to fit a CoG with $b=16.1^{+7.4}_{-3.9}$ km s$^{-1}$,
shown in Figure 9,  to the four
observed \ion{Fe}{2}\ lines.  
Resulting column densities, abundances, and metallicities 
for the ion species are listed in Table 8.  We measure an \ion{Fe}{2}\
metallicity of [\ion{Fe}{2}/\ion{H}{1}]$_{\rm C}=$$-0.42^{+0.15}_{-0.12}$.  
Our \ion{N}{1}\ measurement from $\lambda$1199.55 of 
[\ion{N}{1}/\ion{H}{1}]$_{\rm C}=$$-1.09^{+0.16}_{-0.15}$ is greater 
than the G01 
measurement of [\ion{N}{1}/\ion{H}{1}]$_{\rm C}=$$-1.28\pm0.09$ from the 
same data.  The data quality 
in the SiC1b channel is insufficient to adequately measure
\ion{O}{1}\ absorption, though we are able to place a 3$\sigma$ 
upper limit on the metallicity based on [\ion{O}{1}/\ion{H}{1}]$_{\rm C}$
at $<0.5$ solar.  However, we note that we are able to measure the
\ion{O}{1}\ $\lambda\lambda$929.52, 936.63 lines at the 2$\sigma$ level, 
indicating that the metallicity along this sightline may be relatively high.
The absorption profile of the 929.52 \AA\ line strongly hints at the 
two-component structure seen in \ion{H}{1}.  
Higher quality {\it FUSE} data especially for the SiC channels, 
along with STIS 
observations over a larger bandpass 
for this target, would allow a far more complete investigation
of metallicities along this sightline. 

In an attempt to investigate 
the sub-component structure of Complex C absorption
in this sightline, we have performed seperate abundance analyses of the 
components centered at $V_{LSR}\sim-175$ km s$^{-1}$ (integration
range $-210\leq V_{LSR}\leq-150$ km s$^{-1}$) and $V_{LSR}\sim-130$ km s$^{-1}$
(integration range $-150\leq V_{LSR}\leq-95$ 
km s$^{-1}$).  This is similar to the approach taken by Murphy et al. (2000) 
where these components were fitted separately.  Results are presented in
Table 8.  Measurements of 
[\ion{Fe}{2}/\ion{H}{1}]$_{\rm C}$ for the $-175$ and $-130$ km s$^{-1}$
components are $-0.71^{+0.14}_{-0.15}$ and $-0.35^{+0.10}_{-0.09}$,
respectively, suggesting that the lower velocity component may be of a 
significantly higher metallicity than the higher velocity component.  
If the lower velocity component is indeed at a smaller distance, then it
is possible that it could have a higher degree of enrichment from 
the Galaxy.
In addition, we are able to make tenuous measurements of the 
\ion{O}{1}\ $\lambda\lambda$929.52, 936.63 lines in the lower-velocity 
component at $W_{\lambda}=$ 48$\pm$16 and 53$\pm$19, respectively, consistent
with a metallicity near 0.3 solar.  This value, however, is highly uncertain,
and better quality data are necessary for confirmation.  

\subsection{PG 1259+593}

Of these eight 
sightlines, the data for PG 1259+593 are of the highest quality. 
The {\it FUSE} data have an unbinned S/N of $\sim8$ in the 1040-1050 \AA\ range
of the LiF1a channel, while the STIS spectrum covers useful 
absorption lines in the 1150-1700 \AA\ range.  The bulk of these data were
recently analyzed by Richter et al. (2001a) who measured a metallicity
from the \ion{O}{1}\ lines of $\sim0.1$ solar.  The data presented here 
include an additional 400 ks of exposure time obtained with {\it FUSE},
which allowed us to refine these abundances and obtain a new measurement
of \ion{Ar}{1}.  

Figure 10 shows some of the observed absorption profiles along with the LDS 
profile of \ion{H}{1}\ emission.  The \ion{H}{1}\ profile shows a clear 
division between the Complex C high-velocity component, the 
intermediate-velocity component (the IV Arch), and the Galactic component. The
integration range is thus set at $-155\leq V_{LSR}\leq-95$ km s$^{-1}$, 
with little chance that profiles are contaminated by
intermediate-velocity gas.
The metallicity of the IV Arch is considered by Richter et al. (2001a),
who find solar abundances and conclude the cloud to be of 
Galactic origin.  Measured values of $W_{\lambda}$ are shown in Table 9.

\placetable{t9}

We have measured six \ion{O}{1}, eight \ion{Fe}{2}, 
four \ion{Si}{2}, two \ion{S}{2},
and two \ion{Si}{2}\ lines to which we have fitted a CoG with 
$b=10.0^{+1.9}_{-1.5}$ km s$^{-1}$, shown in Figure 11.  Richter et
al. (2001a) empirically determined a similar value,  
$b=9.8^{+4.7}_{-1.2}$ km s$^{-1}$.  
Resulting column densities, abundances, and 
metallicities for the ion species are listed in Table 9.  The Complex C
abundances measured for this sightline are among the lowest of the sightlines
in this sample.  We confirm the result of Richter et al. (2001a) for 
\ion{O}{1}\, where we measure  
[\ion{O}{1}/\ion{H}{1}]$_{\rm C}=$$-1.00^{+0.19}_{-0.25}$, 
implying a metallicity of
$\sim0.1$ solar.  The \ion{S}{2}\ abundance is also low, 
[\ion{S}{2}/\ion{H}{1}]$_{\rm C}=$$-0.74^{+0.13}_{-0.13}$,
with only the 
Mrk 290 sightline indicating a lower \ion{S}{2}\ metallicity. The high quality
of these new data allow a measurement of 
[\ion{Ar}{1}/\ion{H}{1}]$_{\rm C}=$$-1.22^{+0.13}_{-0.16}$. 
This target is the only sightline for
which a $>3\sigma$ detection can be made from 
\ion{Ar}{1}\ $\lambda$1048.22.

\subsection{PG 1351+640}

Currently available {\it FUSE} data for this sightline are of fair quality, 
with
an average S/N of $\sim3$ in the 1040-1050 \AA\ range for the LiF1a channel  
per unbinned pixel.  However, the target was not observed by the SiC1 channels
thus limiting our ability to detect absorption at $\lambda\lesssim990$ \AA.  
The STIS data are of good quality, although
the spectral range of the observations (1195-1299 \AA) 
is more limited than the STIS data for PG 1259+593.  

The observed absorption profiles, along with the LDS profile of \ion{H}{1}\ 
emission, are shown in Figure 12.  The high-velocity \ion{H}{1}\ component of
the profile appears to be separated from the  prominent 
intermediate-velocity feature, and we choose the integration range of 
Complex C to span the extent of the high-velocity emission 
($-190\leq V_{LSR}\leq-95$ km s$^{-1}$).  Measured values of $W_{\lambda}$ are
shown in Table 10.  The \ion{O}{1}\ $\lambda$1039.230 Complex C feature is
contaminated by Galactic H$_{2}$ (5-0) R(2) Lyman 
absorption, and thus only an upper
limit to the Complex C contribution can be established.  
In addition, poor data quality
in the SiC channels precludes any determination of \ion{O}{1}\
abundance based on the 920-950 \AA\ lines.  
Further {\it FUSE} observations would obviously be very helpful.
In the STIS spectrum, strong absorption features of unknown origin
coincide with the Complex C components of \ion{N}{1}\ $\lambda$1199.55 and 
\ion{S}{2}\ $\lambda$1259.52, although we are able to measure the \ion{S}{2}\
$\lambda$1253.80 line.  

\placetable{t10}

We have used the two detected \ion{Fe}{2}\ lines to determine a best-fit CoG 
with $b=13.7^{+3.3}_{-2.5}$ km s$^{-1}$.  Resulting column 
densities, abundances, and metallicities for the ion species are listed 
in Table 10.  We measure an abundance 
[\ion{Fe}{2}/\ion{H}{1}]$_{\rm C}=$$-0.45^{+0.26}_{-0.17}$, 
which is among the highest of
the sightlines in this sample.  The upper limit placed on the metallicity
based on the \ion{O}{1}\ $\lambda$1039.23/H$_{2}$ blend is at $<0.5$ solar.  
Limits on N(\ion{O}{1}) based on lines in the SiC channels do not improve
on this value.
We expect a value of $W_{\lambda}$ in
the range 100-130 m\AA, corresponding to a metallicity range of 0.1-0.25 
solar, for the \ion{O}{1}\ $\lambda948.69$ line.
The upper limit 
[\ion{S}{2}/\ion{H}{1}]$_{\rm C}<$$-0.49$ is consistent with the trend 
we observe
in this sample for [\ion{S}{2}/\ion{H}{1}]$_{\rm C}$ versus $N$(\ion{H}{1}). 
 
\subsection{PG 1626+554}

The {\it FUSE} data for this final target in our sample are poor, 
with an average
S/N of $\sim1.2$ per unbinned pixel in the 1040-1050 \AA\ range.  The data
consists of only 8.3 ks of exposure time, the shortest exposure in this sample.
In addition, GHRS or STIS data for this target do not exist.   
Observed 
absorption profiles are shown in Figure 14 along with the LDS profile of 
\ion{H}{1}\ emission.  The \ion{H}{1}\ profile shows a prominent high-velocity
component, well-separated from lower-velocity emission, spanning the
velocity range of $-155\leq V_{LSR}\leq-75$ km s$^{-1}$.  This target
is not included in the Wakker et al. (2001) Effelsberg survey, so we 
use the LDS
data to measure a neutral hydrogen column of the high-velocity component, 
$N$(\ion{H}{1})$=(6.2\pm0.4)\times10^{19}$ cm$^{-2}$.  We note that the 
systematic error associated with abundance measurements in this sightline are
quite high due to the large beam size for the $N$(\ion{H}{1}) measurement.  

Owing to data quality, we are unable to make a 3$\sigma$ detection of any of
the Complex C absorption features; the most significant detection is the 
\ion{Fe}{2}\ $\lambda$1144.94 line at the 2$\sigma$ level.
The \ion{O}{1}\ $\lambda$1039.23 feature is prominent, but 
strong Galactic H$_{2}$ contamination is likely. 
Given the possible H$_{2}$ contamination, it would be useful to obtain
high-quality SiC data to analyze \ion{O}{1}\ lines in the 920-950 \AA\ region. 
The \ion{O}{1}\ $\lambda948.686$ line is expected to have a $W_{\lambda}$ 
value in the range 80-100 m\AA, corresponding to a range in metallicity
of 0.1-0.25 solar.
Measurements of equivalent width, column density, abundance, and metallicity
are summarized in Table 11.  Column densities were determined as for Mrk 290. 

\placetable{t11}

The upper limit of [\ion{O}{1}/\ion{H}{1}]$_{\rm C}<$$-0.79$ is quite low and,
aside from the uncertain $N$(\ion{H}{1}) from the LDS data,
suggests that the metallicity may be close to that of the $\sim0.1$ solar value
along the PG 1259+593 sightline.  The \ion{Fe}{2}\ upper limit 
is also quite low, and the 2$\sigma$ measurement of the 1144.939 \AA\ line at
$W_{\lambda}$= 81$\pm$39 suggests an \ion{Fe}{2}\ abundance below 0.1 solar.   
Clearly this 
is an interesting sightline for further {\it FUSE} and {\it HST} 
observations, along with higher-resolution \ion{H}{1}\ data.

\section{Discussion}

\subsection{Ionization Effects on Elemental Abundances}

For the eight sightlines, we have calculated abundances of various ion species,
\ion{O}{1}, \ion{S}{2}, \ion{Fe}{2}, \ion{Si}{2}, and \ion{N}{1}, which 
are the dominant ionization states in warm neutral clouds.  However,
there are certain to be regions in these clouds with ionized hydrogen.  As a 
result, the measured column of \ion{H}{1}\ does not fully reflect the total
column of H along the line of sight.  This is not a problem for elements such 
as N or O, where the ionization state is coupled to that of H through
charge-exchange interactions.  Elements such as S and Si, however,
are certain to be singly-ionized in both \ion{H}{1}\ and \ion{H}{2}\ 
regions.  Therefore, in order to make definitive statements regarding 
{\it elemental} abundances, one must take into account possible photoionization
corrections to ion abundances.

In Figure 15, we have plotted the runs of \ion{S}{2}, \ion{Si}{2}, \ion{Fe}{2},
and \ion{O}{1}\ abundance versus $N$(\ion{H}{1}).  The abundance 
[\ion{S}{2}/\ion{H}{1}], and to a lesser extent [\ion{Si}{2}/\ion{H}{1}], 
shows a trend of decreasing abundance with column density.  If one assumes 
that the elemental abundance does not vary significantly as a function of 
column density, then such a trend can be explained by an ionization effect.

In order to model the effect of column density on ion abundances, 
we have generated a grid of photoionization
models using the code CLOUDY (Ferland 1996).  We make the simplified assumption
that the absorbing gas can be treated  as plane-parallel slabs illuminated 
by incident radiation dominated by OB associations.  The models assume a 
$T=35,000$ K Kurucz model atmosphere and a 
cloud metallicity of $Z_{\sun}=0.1$. 
While the extent of the incident ionizing field on Complex C is not
well known, H$\alpha$ emission measures of Complex C have been argued
(Bland-Hawthorn \& Putman 2001; Weiner et al. 2002)
to be consistent with radiation from our Galaxy at the level of
log $\phi \approx 5.5$ (photons cm$^{-2}$ s$^{-1}$),
which we assume as the normally-incident ionizing 
flux in the models.  For three assumed gas volume densities, $n_{\rm H}$, 
we calculate the logarithmic difference between [X$^{ion}$/\ion{H}{1}] and 
[X$^{element}$/H] versus $N$(\ion{H}{1}), shown in Figure 16 for S, Si, and Fe.
We find that the ionization corrections necessary for O and N are negligible, 
and that [\ion{O}{1}/\ion{H}{1}]$_{\rm C}$ and 
[\ion{N}{1}/\ion{H}{1}]$_{\rm C}$ should reflect actual values of [O/H] and 
[N/H].  It would be useful to check this for N by measuring Complex C
\ion{N}{2}\ $\lambda1084$. However, this absorption feature cannot be measured
in these sightlines since it is saturated and blended with saturated 
Galactic absorption.

Given the simplified nature of these models and the uncertainty in the 
actual gas densities, these models are for illustrative purposes only and
are designed mainly to explain the general trend observed in the ion 
abundances versus \ion{H}{1} column.  The models show that the spread in values
of [\ion{S}{2}/\ion{H}{1}]$_{\rm C}$ observed in these sightlines is more
a reflection of ionization conditions in the cloud than actual variations
in [S/H]. A similar statement can be made for the [\ion{Si}{2}] trend.
For the intermediate density model ($n_{\rm H}=0.03$ cm$^{-3}$),
the ionization correction to the [\ion{S}{2}/\ion{H}{1}]$_{\rm C}$
puts those  values more in line with the 
[\ion{O}{1}/\ion{H}{1}]$_{\rm C}$ measurements. 
Fe is not subject to as large a photoionization
correction, which may explain the lack of a 
strong trend in [\ion{Fe}{2}/\ion{H}{1}]$_{\rm C}$ versus $N$(\ion{H}{1}).

\subsection{The Origin of Complex C}

In Table 12, we show the measured abundances, relative to solar, of 
\ion{O}{1}, \ion{S}{2}, \ion{Fe}{2}, \ion{Si}{2}, and \ion{N}{1}.  
As demonstrated in \S\ 4.1, 
many of these ion species' abundances are subject to 
ionization corrections in order to convert to elemental abundances.  Oxygen and
nitrogen are particularly insensitive to such effects, and 
the values listed should accurately reflect
the true values of [O/H]$_{\rm C}$ and [N/H]$_{\rm C}$.  

\placetable{t12}

We find that the 
metallicity of Complex C ranges from 0.1-0.25 solar for the three sightlines 
in which \ion{O}{1}\ could be measured.  The upper limits on 
[\ion{O}{1}/\ion{H}{1}]$_{\rm C}$ established for the other sightlines are 
consistent with such a range.   As stated by G01, the moderate abundances 
imply that Complex C is unlikely to be representative of purely
infalling extragalactic gas.  
Further, we find that the metallicity varies from 
sightline to sightline; the [\ion{O}{1}/\ion{H}{1}]$_{\rm C}$ and
ionization-corrected [\ion{S}{2}/\ion{H}{1}]$_{\rm C}$, within their
$1\sigma$ error bars, suggest variations by factors of 2-3.  
Although N is produced through 
different mechanisms than O, the two cases for which \ion{N}{1}\ could be 
measured differ in abundance by 0.7 dex, further suggesting that Complex
C cannot be characterized by a single metallicity.  
Metallicities in
the range 0.1-0.3 solar are characteristic of tidally-disrupted
dwarf-satellites (LMC and SMC) or gas in the outer disk ($>1.5R_{\sun}$) 
of the Galaxy (Gibson 2002). While the calculated 
metallicities are still somewhat low to assign a Galactic origin,
it is tempting to assign a mixed origin to Complex C,
with gas of Galactic (and/or satellite) origin blending with 
infalling low-metallicity clouds.  
If this is the case, one might expect, with better data, to see a trend of
lower metallicity with higher $N$(\ion{H}{1}), since the mixing of infalling
gas with enriched Galactic disk gas may take considerable time.  The
interpretation of Complex C metallicity may therefore depend on the 
alignment of the background AGN with the cloud core or halo.
We suggest that further {\it FUSE} and {\it HST} observations 
of many of these sightlines, together with higher-resolution 21-cm 
\ion{H}{1}\ maps,
will more completely address these issues. 

In order to further explore issues concerning the origin of Complex C, we have
calculated elemental 
abundances relative to \ion{O}{1}.  We consider data only from the 
three sightlines (Mrk 279, Mrk 817, and PG 1259+593) for which an \ion{O}{1}\
column density could be measured.  Figure 17 shows measured values of
[\ion{S}{2}/\ion{O}{1}], [\ion{Si}{2}/\ion{O}{1}], [\ion{Fe}{2}/\ion{O}{1}], 
and [\ion{N}{1}/\ion{O}{1}], where we have plotted the column density weighted
mean values for the 
three sightlines.  The upper limit for [\ion{N}{1}/\ion{O}{1}] is taken as
the least restrictive of the limits from the various sightlines. 

Using the ionization corrections discussed in \S\ 4.1 for the intermediate
density model ($n_{\rm H}=0.03$ cm$^{-3}$) at log $N$(\ion{H}{1})=19.5, 
we can determine relative elemental abundances by adding an offset to
the values illustrated in Figure 17.  With these corrections, we find
that [S/O]$_{\rm C}$ and [Si/O]$_{\rm C}$ are consistent with a solar 
relative abundance.  The fact that Si is not depleted implies that dust is
not a significant constituent of Complex C.  The possible absence of dust 
in Complex C is reinforced by the non-detection of H$_{2}$, whose
formation is facilitated by surfaces of interstellar dust grains 
(Shull \& Beckwith 1982).   
The corrected [Fe/O] is sub-solar
by $\sim0.2$ dex, which in the absence of dust depletions  
suggests a slight enhancement of the $\alpha$-elements O, Si, and S.  
We note that [\ion{Fe}{2}/\ion{S}{2}] and [\ion{Fe}{2}/\ion{Si}{2}] are 
sub-solar. As a result, we can be reasonably confident in the corrected
[Fe/O] given appropriate corrections to \ion{S}{2}\ and \ion{Si}{2}. 
In addition, N is depleted by at least 0.5 dex relative to O for these 
three sightlines.  Evidence for under-abundant N is also 
apparent for the Mrk 876 sightline, where \ion{N}{1} is reduced by 
0.8 dex from solar with respect to the $\alpha$-element ion \ion{Si}{2}, much
greater than any possible ionization correction along this sightline.

The enrichment of $\alpha$-elements relative to Fe and the significant
depletion of N suggest that the metals were produced primarily by massive stars
and injected into the ISM by Type II SNe.  The gas could have then been 
removed from the star-forming regions before Fe and N production and 
delivery could pollute the ISM.  
If the gas originated in a star-forming region
with sufficiently correlated SN activity, then it could have been driven 
into the halo through chimneys (Norman \& Ikeuchi 1989) 
in a ``Galactic fountain'' scenario.  If the gas 
originated in the inner disk, then additional mixing with infalling 
extragalactic gas would be necessary to explain the metallicities.  
Alternatively, the similarity in metallicity to the LMC and SMC suggests
that the gas may have been tidally stripped from the outer disk of the
Milky Way or an unknown dwarf irregular galaxy.  
These scenarios are all speculative.  Our study suggests, however,
that the data are not consistent with a simple model of an infalling
low-metallicity cloud of extragalactic origin.

\acknowledgments

This work is based upon data obtained for the Guaranteed Time Team by the 
NASA-CNES-CSA {\it FUSE} mission operated by the Johns Hopkins University.  
Financial
Support to US participants has been provided by NASA contract NAS5-32985.  
Financial support has also 
been provided by the NASA Theory program at the University of Colorado
through grant
NAG5-7262.  Additional support has been provided through grant 
GO-06593-01-A from the Space Telescope Science Institute.  We express our
thanks to Bart Wakker and Brad Gibson for useful discussions 
regarding this work.  
Thanks also to Bart Wakker for providing the Effelsberg \ion{H}{1}\
data for Mrk 279 and Mrk 876.

\begin{figure}
\figurenum{1}
\plotone{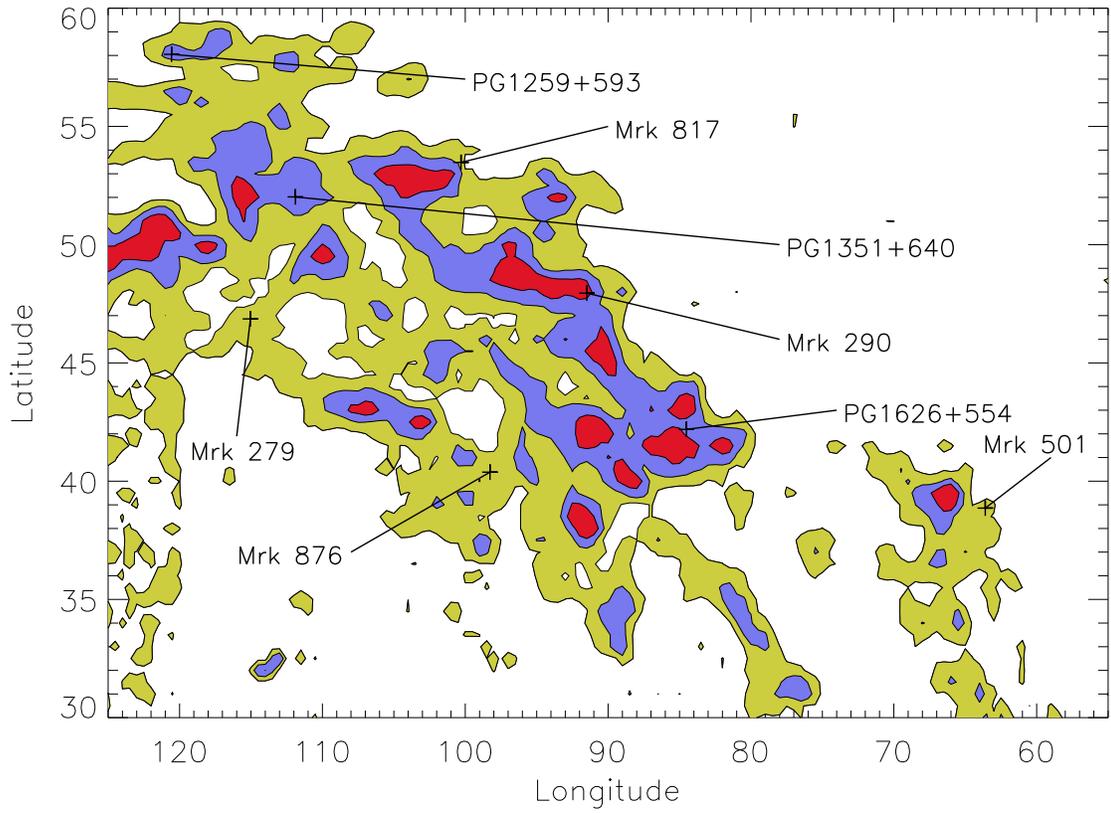}
\caption{Contours of \ion{H}{1}\ column density from the Leiden-Dwingeloo 
Survey (Hartmann \& Burton 1997) for the high-velocity gas of Complex C 
($-210\leq V_{LSR}\leq-95$ km s$^{-1}$). Contour levels are 
$N$(\ion{H}{1})$=1, 3$, and $6\times10^{19}$ cm$^{-2}$. 
The locations of the eight sightlines for this survey are labeled.}
\end{figure}

\begin{figure}
\figurenum{2}
\plotone{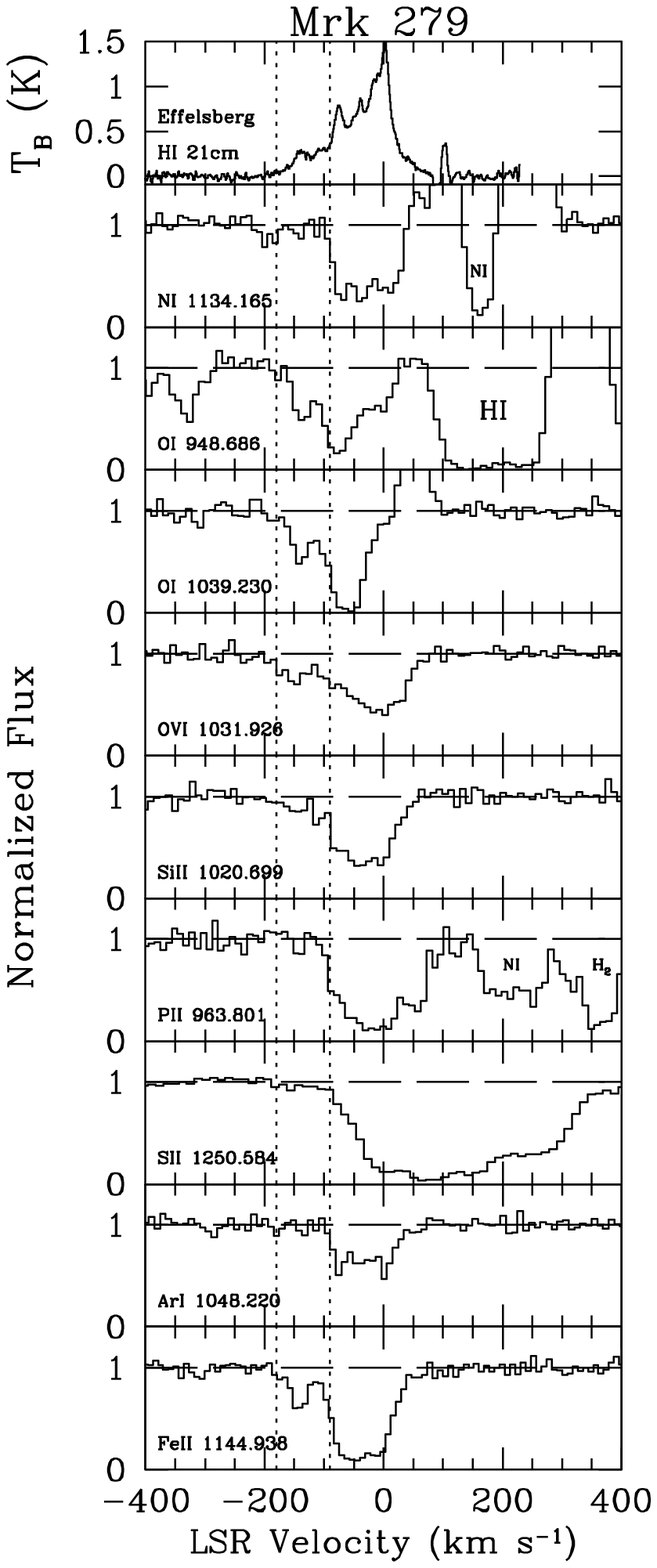}
\caption{A sample of normalized 
absorption profiles from {\it FUSE} and GHRS data for Mrk 279, along with the 
Effelsberg profile of \ion{H}{1} emission (top panel).  The vertical dashed 
lines indicate the $-180$ to $-90$ km s$^{-1}$ range of integration for 
measurements of $W_{\lambda}$.}
\end{figure} 

\begin{figure}
\figurenum{3}
\plotone{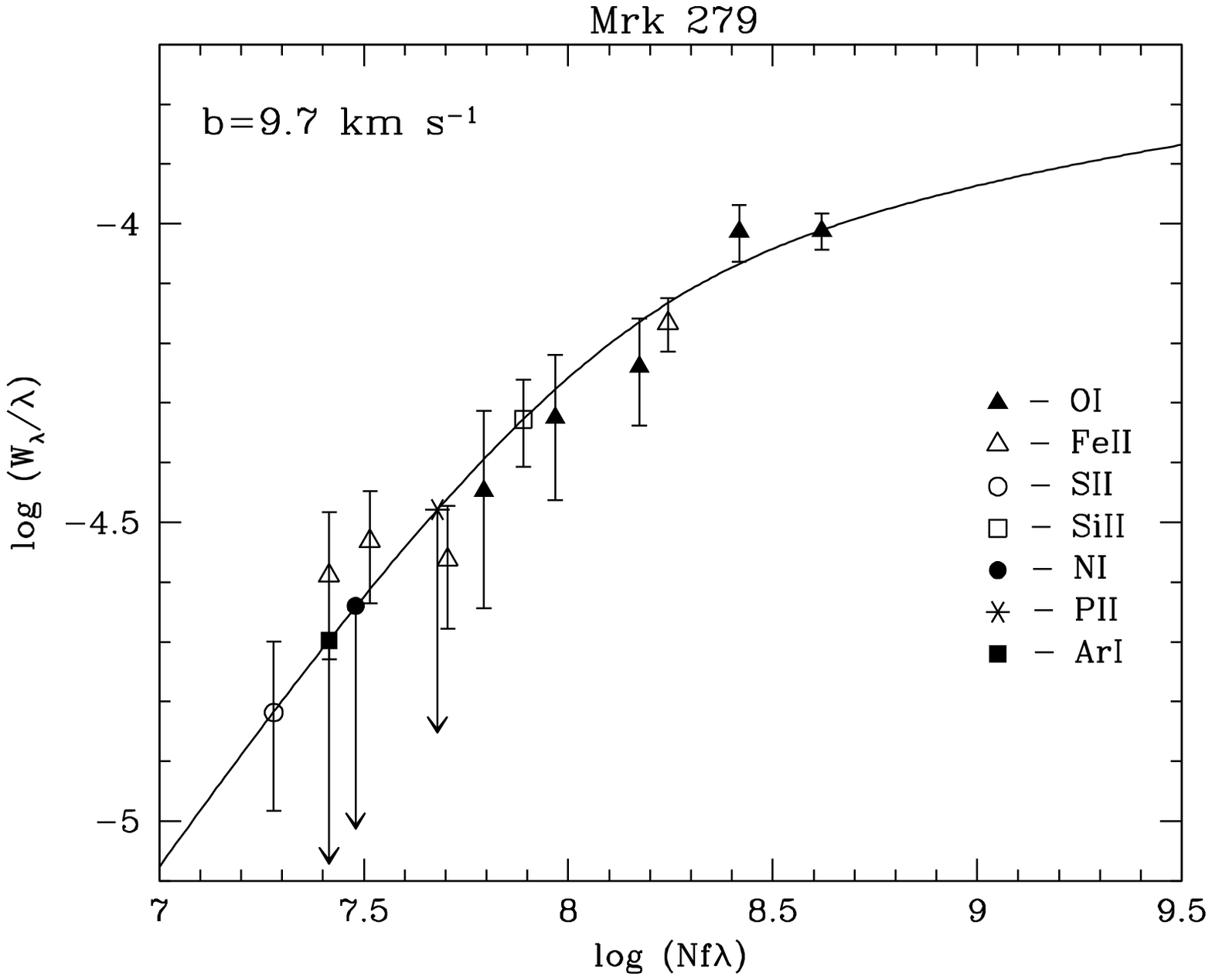}
\caption{Empirical curve of growth for ions observed in the Mrk 279 sightline. The \ion{O}{1}\ and \ion{Fe}{2}\ lines are best fitted by a curve with $b=9.7^{+5.0}_{-2.8}$ km s$^{-1}$.}
\end{figure} 

\begin{figure}
\figurenum{4}
\plotone{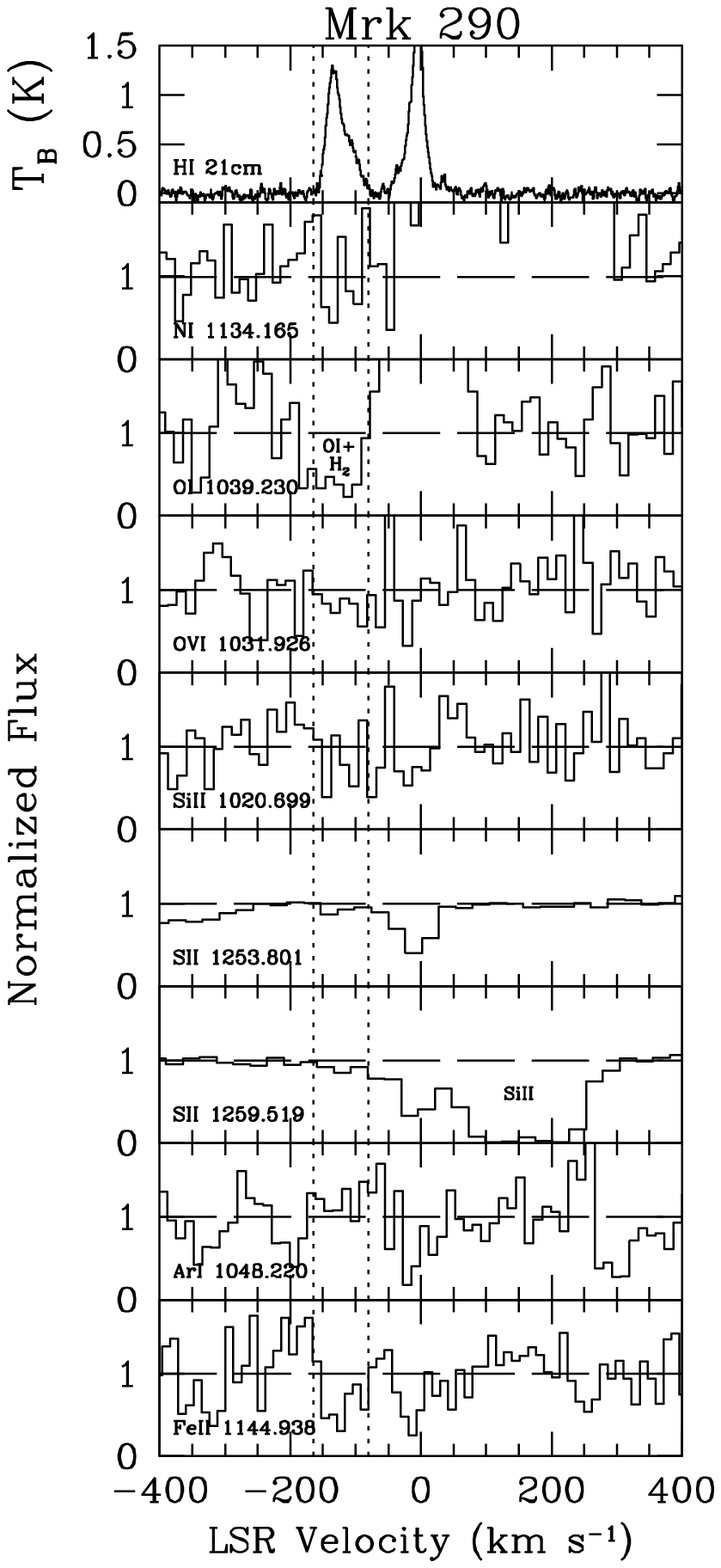}
\caption{A sample of normalized 
absorption profiles from {\it FUSE} and GHRS data for Mrk 290, along with the 
LDS profile of \ion{H}{1} emission (top panel).  The vertical dashed 
lines indicate the $-165$ to $-80$ km s$^{-1}$ range of integration for 
measurements of $W_{\lambda}$.}
\end{figure} 

\begin{figure}
\figurenum{5}
\plotone{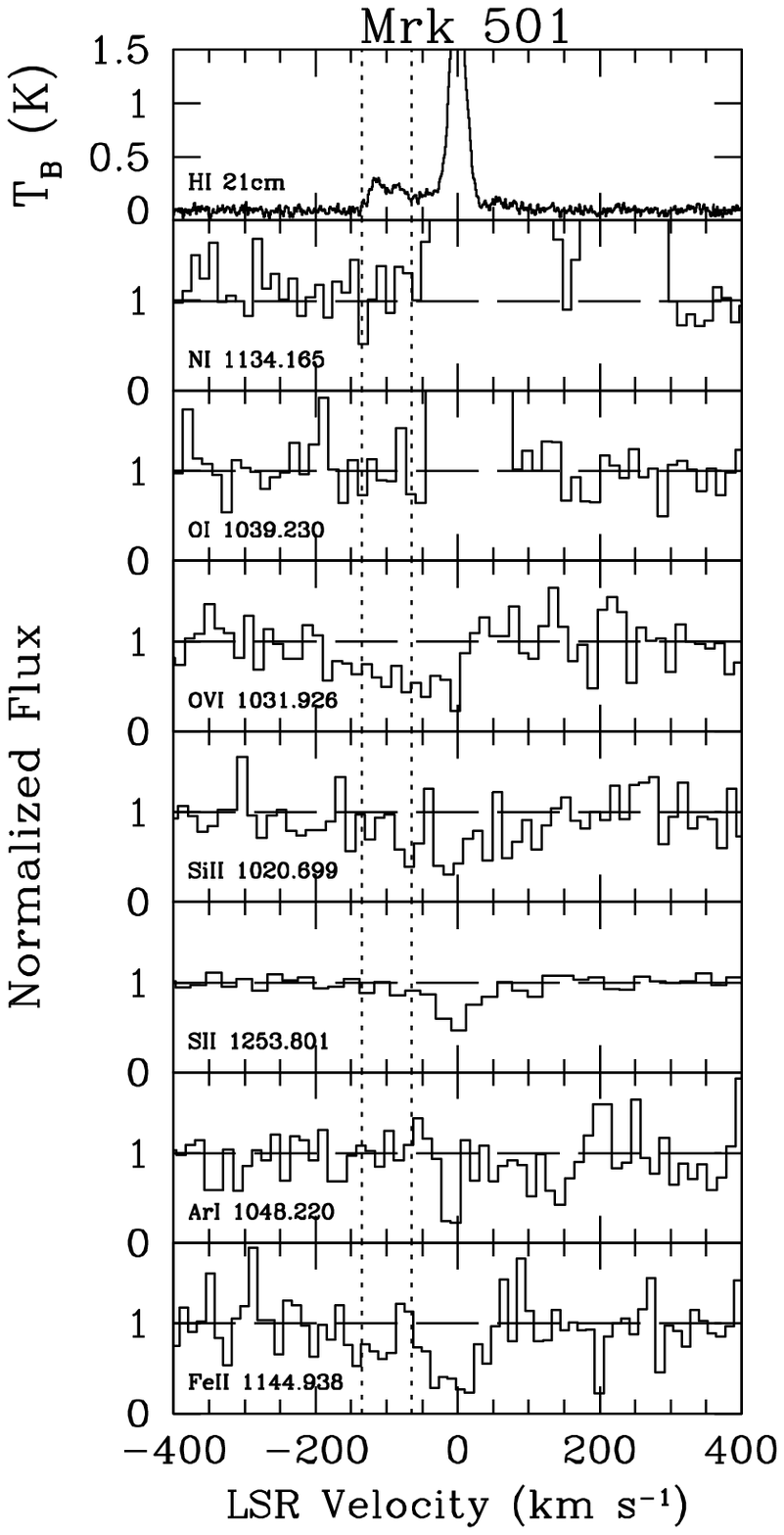}
\caption{A sample of normalized 
absorption profiles from {\it FUSE} and GHRS data for Mrk 501, along with the 
LDS profile of \ion{H}{1} emission (top panel).  The vertical dashed 
lines indicate the $-135$ to $-65$ km s$^{-1}$ range of integration for 
measurements of $W_{\lambda}$.}
\end{figure} 

\begin{figure}
\figurenum{6}
\plotone{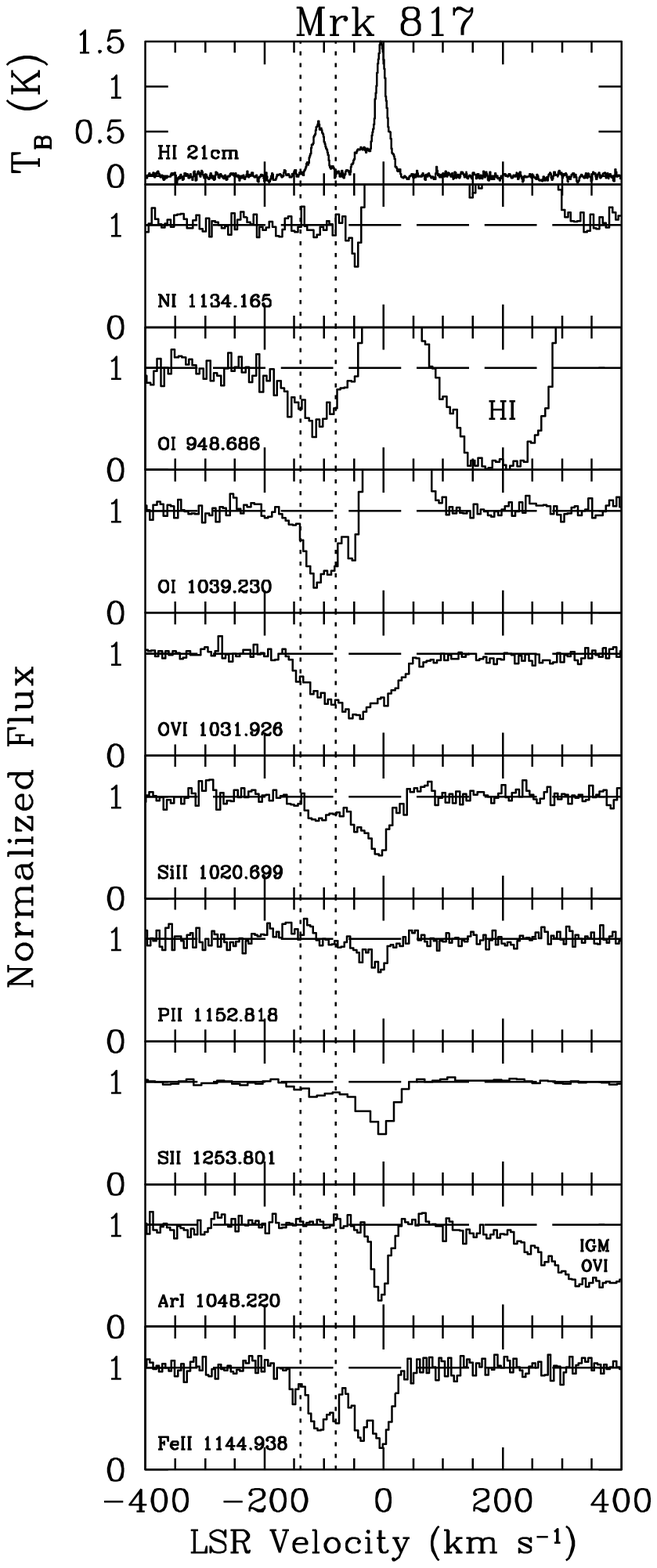}
\caption{A sample of normalized 
absorption profiles from {\it FUSE} and GHRS data for Mrk 817, along with the 
LDS profile of \ion{H}{1} emission (top panel).  The vertical dashed 
lines indicate the $-140$ to $-80$ km s$^{-1}$ range of integration for 
measurements of $W_{\lambda}$.}
\end{figure} 

\begin{figure}
\figurenum{7}
\plotone{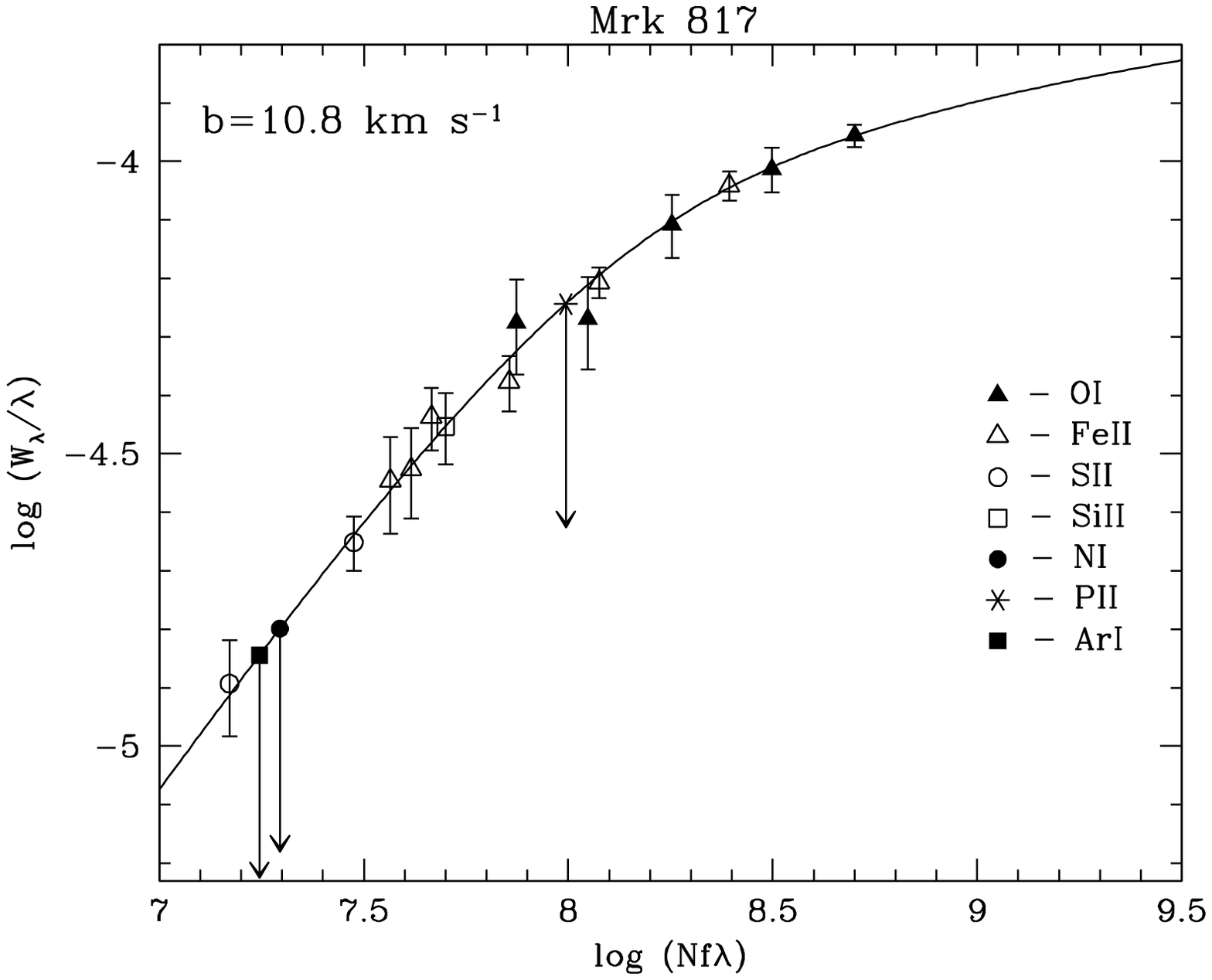}
\caption{Empirical curve of growth for ions observed in the Mrk 817 sightline. The \ion{O}{1}, \ion{Fe}{2}, and \ion{S}{2} lines are best fitted by a curve with $b=10.8^{+3.2}_{-2.2}$ km s$^{-1}$.}
\end{figure} 

\begin{figure}
\figurenum{8}
\plotone{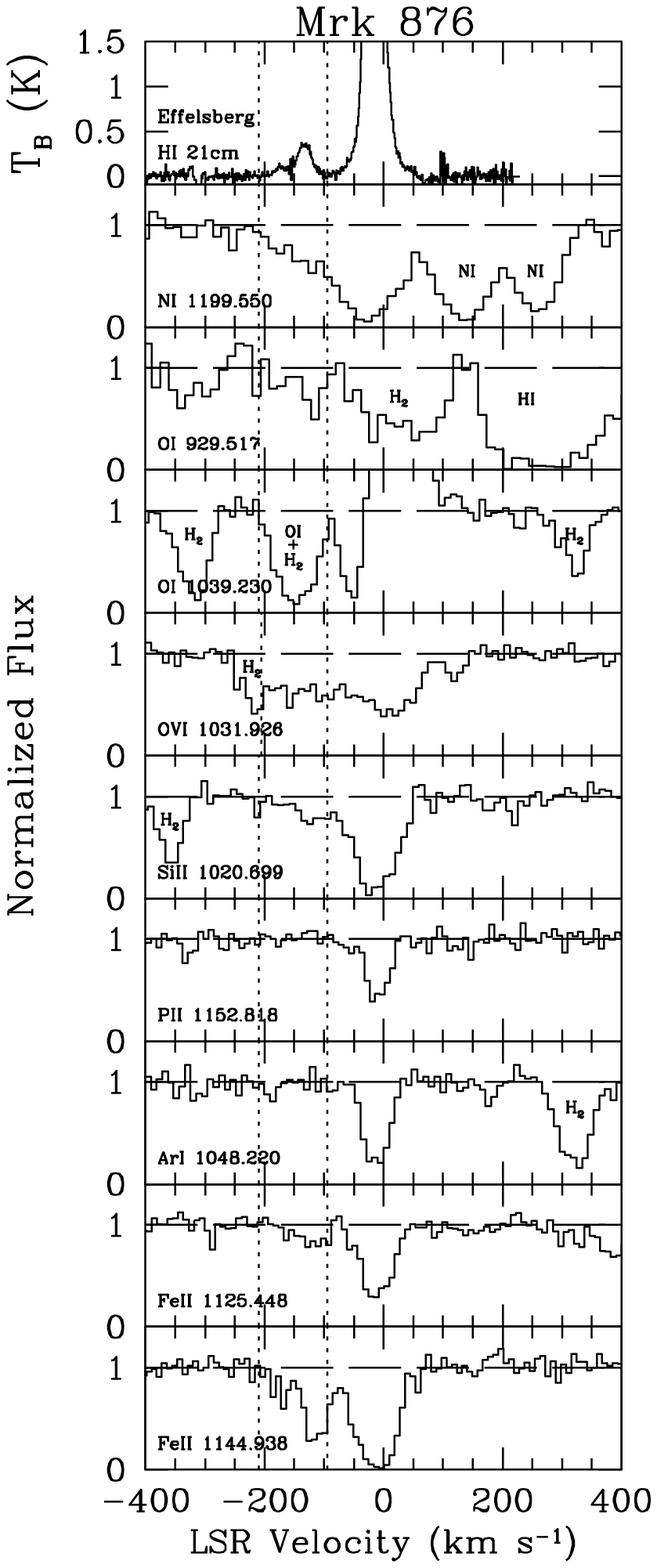}
\caption{A sample of normalized 
absorption profiles from {\it FUSE} and GHRS data for Mrk 876, along with the 
Effelsberg profile of \ion{H}{1} emission (top panel).  The vertical dashed 
lines indicate the $-210$ to $-95$ km s$^{-1}$ range of integration for 
measurements of $W_{\lambda}$.}
\end{figure} 

\begin{figure}
\figurenum{9}
\plotone{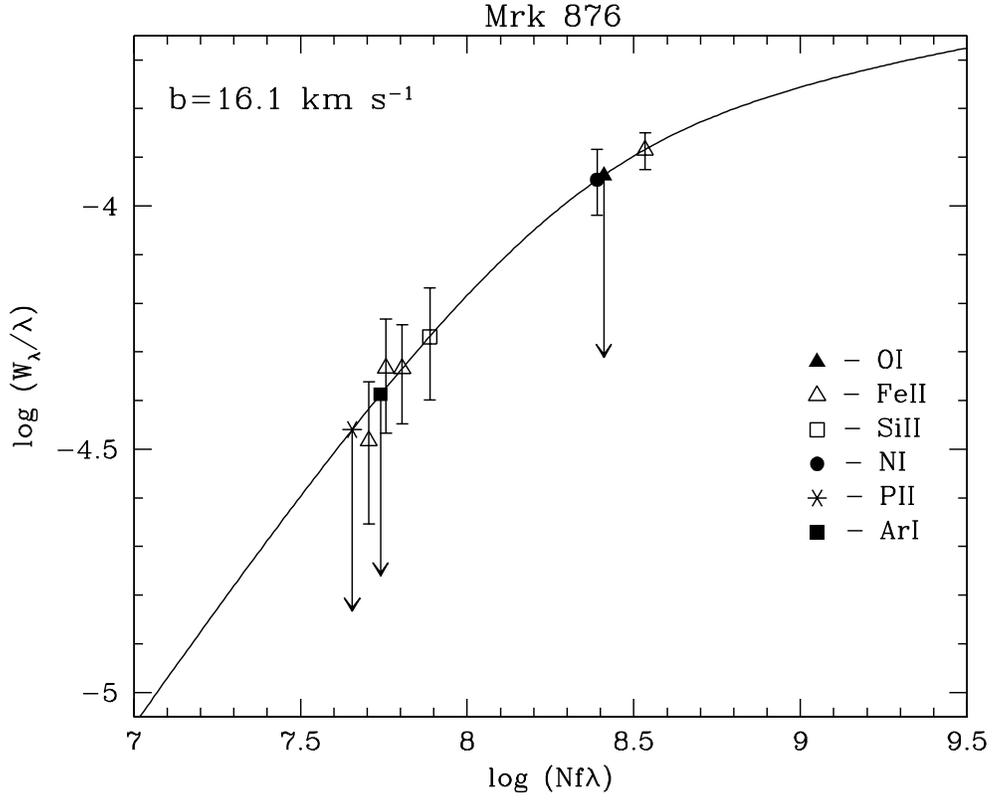}
\caption{Empirical curve of growth for ions observed in the Mrk 876 sightline. The \ion{Fe}{2}\ lines are best fitted by a curve with $b=16.1^{+7.4}_{-3.9}$ km s$^{-1}$.}
\end{figure} 

\begin{figure}
\figurenum{10}
\plotone{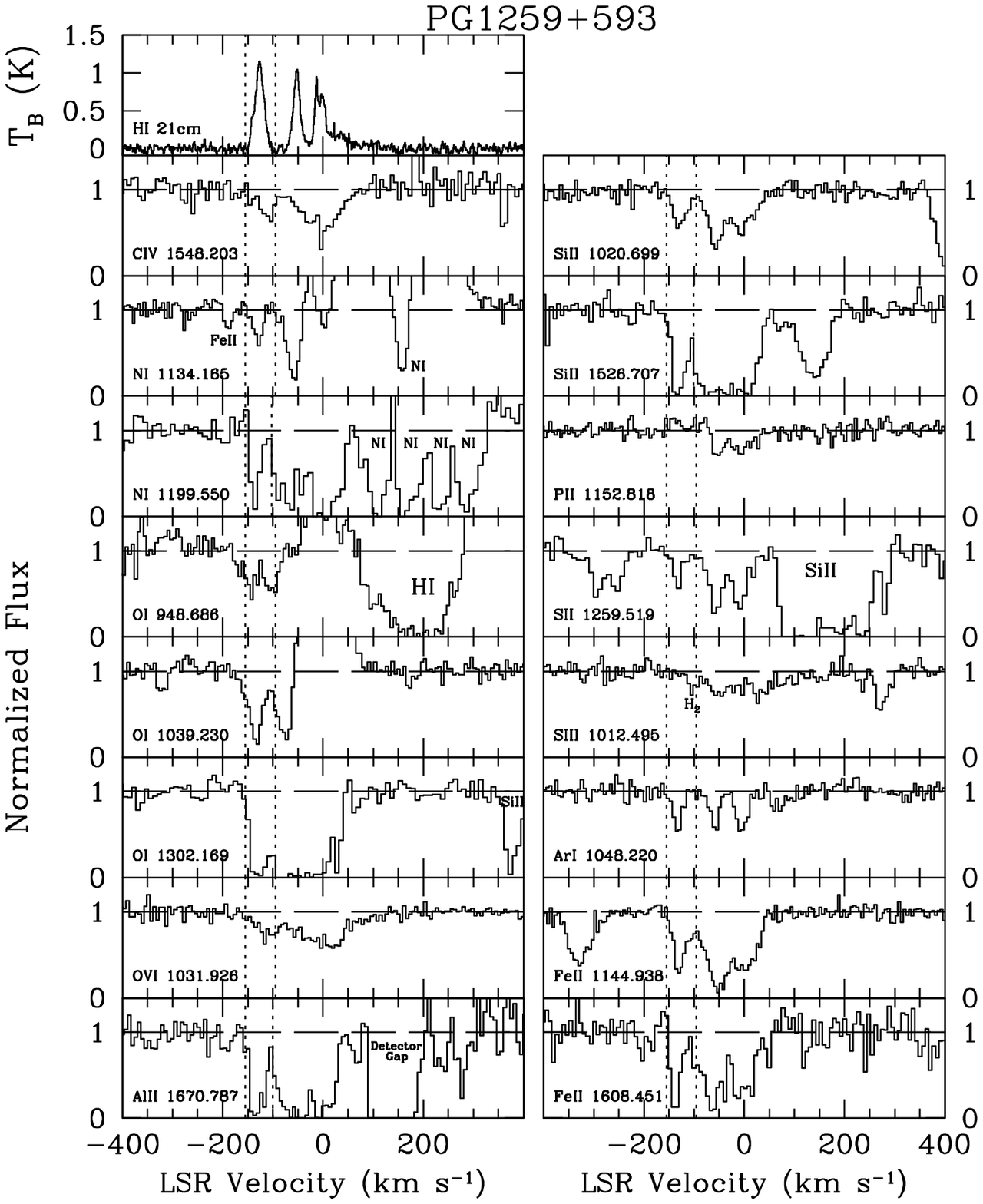}
\caption{A sample of normalized 
absorption profiles from {\it FUSE} and STIS data for PG 1259+593, along with the 
LDS profile of \ion{H}{1} emission (top panel).  The vertical dashed 
lines indicate the $-155$ to $-95$ km s$^{-1}$ range of integration for 
measurements of $W_{\lambda}$.}
\end{figure} 

\begin{figure}
\figurenum{11}
\plotone{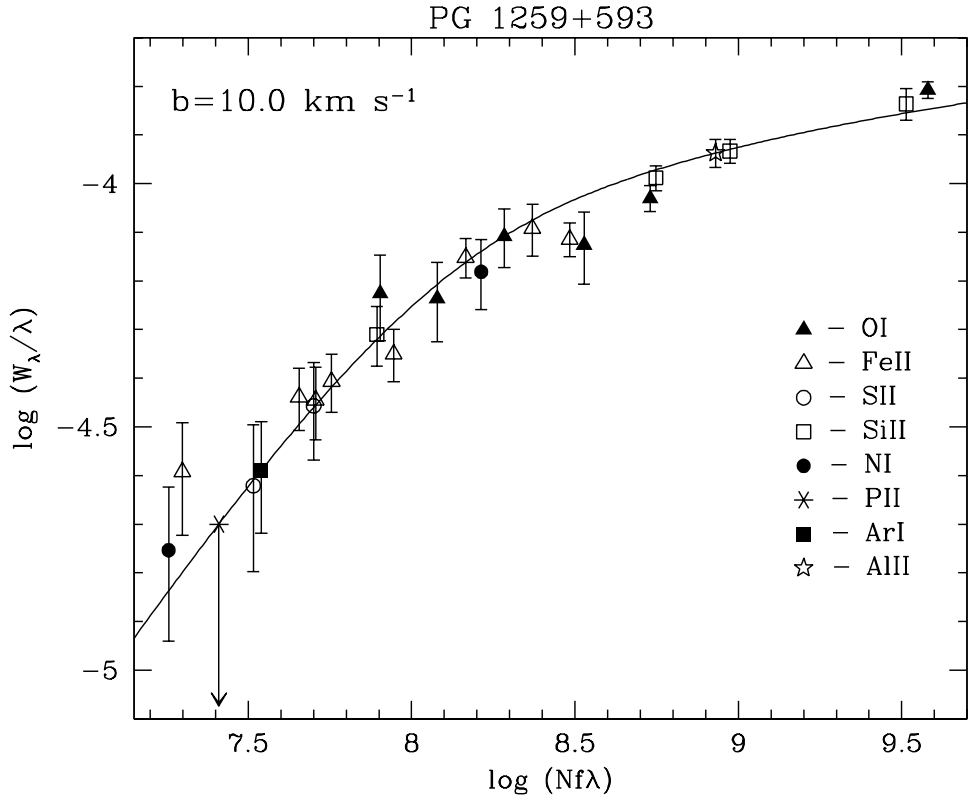}
\caption{Empirical curve of growth for ions observed in the PG 1259+593 sightline. The \ion{O}{1}, \ion{Fe}{2}, \ion{S}{2}, \ion{Si}{2}, and \ion{N}{1}\ lines are best fitted by a curve with $b=10.0^{+1.9}_{-1.5}$ km s$^{-1}$.}
\end{figure} 

\begin{figure}
\figurenum{12}
\plotone{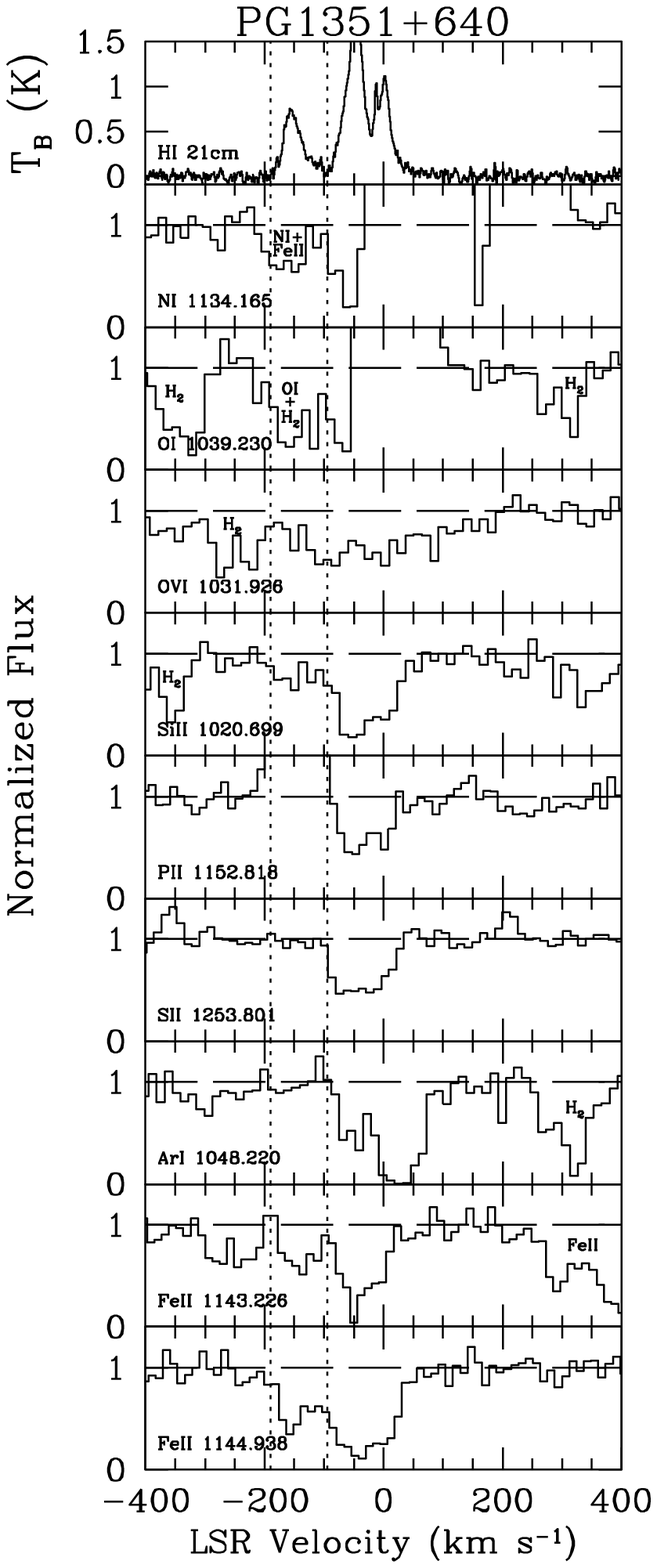}
\caption{A sample of normalized 
absorption profiles from {\it FUSE} and STIS data for PG 1351+640,
along with the 
LDS profile of \ion{H}{1} emission (top panel).  The vertical dashed 
lines indicate the $-190$ to $-95$ km s$^{-1}$ range of integration for 
measurements of $W_{\lambda}$.}
\end{figure} 

\begin{figure}
\figurenum{13}
\plotone{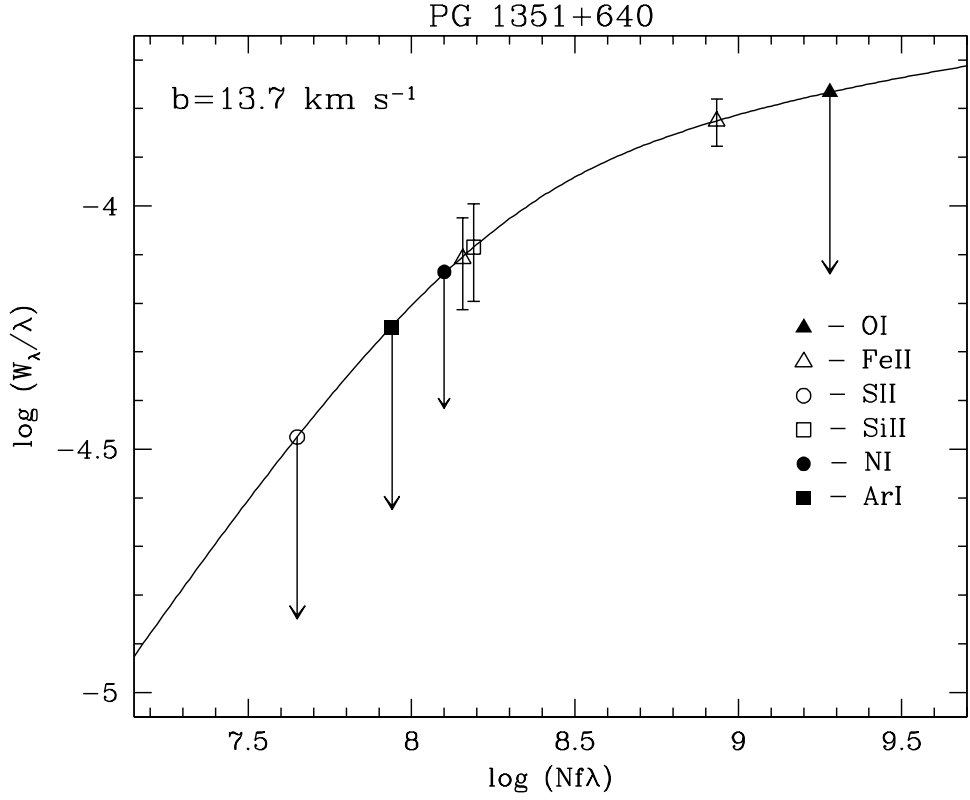}
\caption{Empirical curve of growth for ions observed in the PG 1351+640 sightline. The \ion{Fe}{2}\ lines are best fitted by a curve with $b=13.7^{+3.3}_{-2.5}$ km s$^{-1}$.}
\end{figure}

\begin{figure}
\figurenum{14}
\plotone{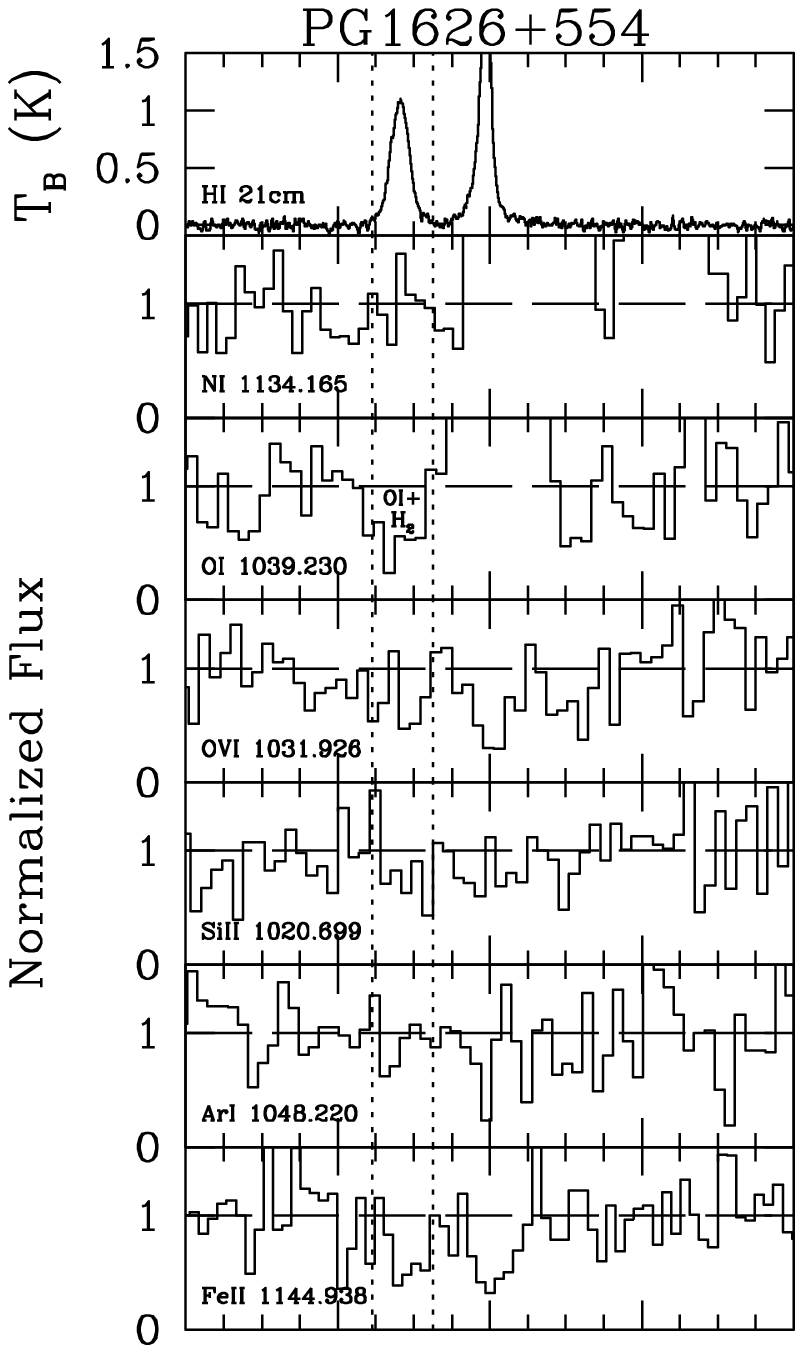}
\caption{A sample of normalized 
absorption profiles from {\it FUSE} data for PG 1626+554, along with the 
LDS profile of \ion{H}{1} emission (top panel).  The vertical dashed 
lines indicate the $-155$ to $-75$ km s$^{-1}$ range of integration for 
measurements of $W_{\lambda}$.}
\end{figure} 

\begin{figure}
\figurenum{15}
\plotone{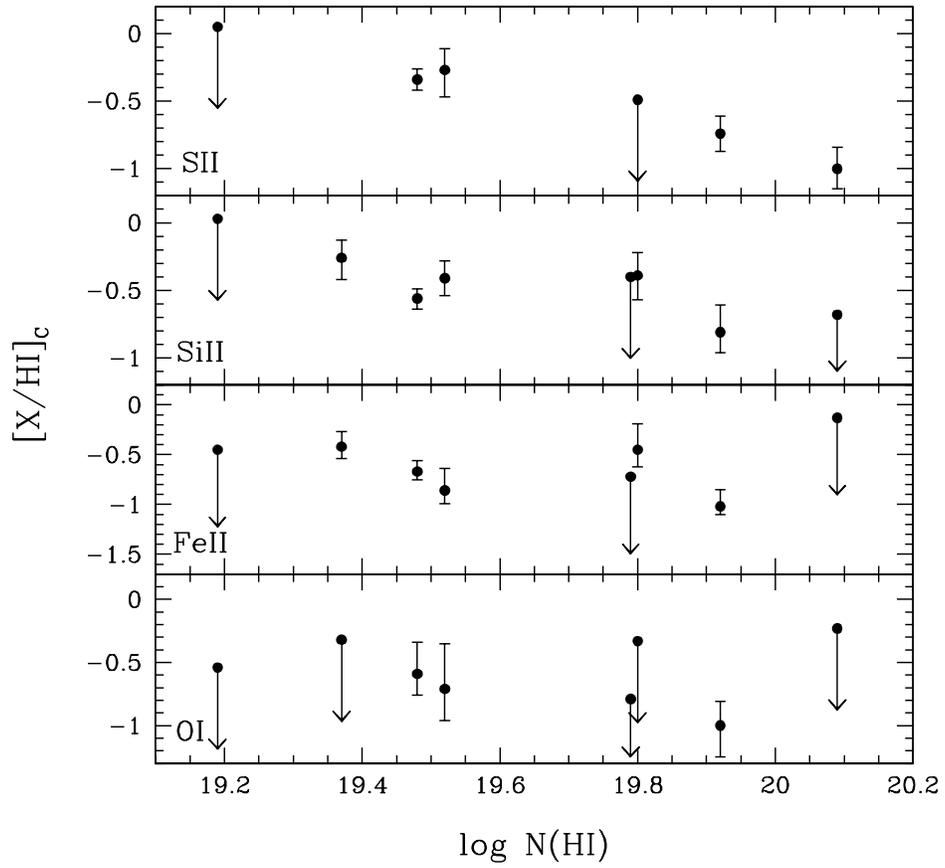}
\caption{Plots of ion abundances of \ion{S}{2}, \ion{Si}{2}, \ion{Fe}{2}, and \ion{O}{1} vs. \ion{H}{1}\ column density for the eight 
sightlines through Complex C.}
\end{figure} 

\begin{figure}
\figurenum{16}
\plotone{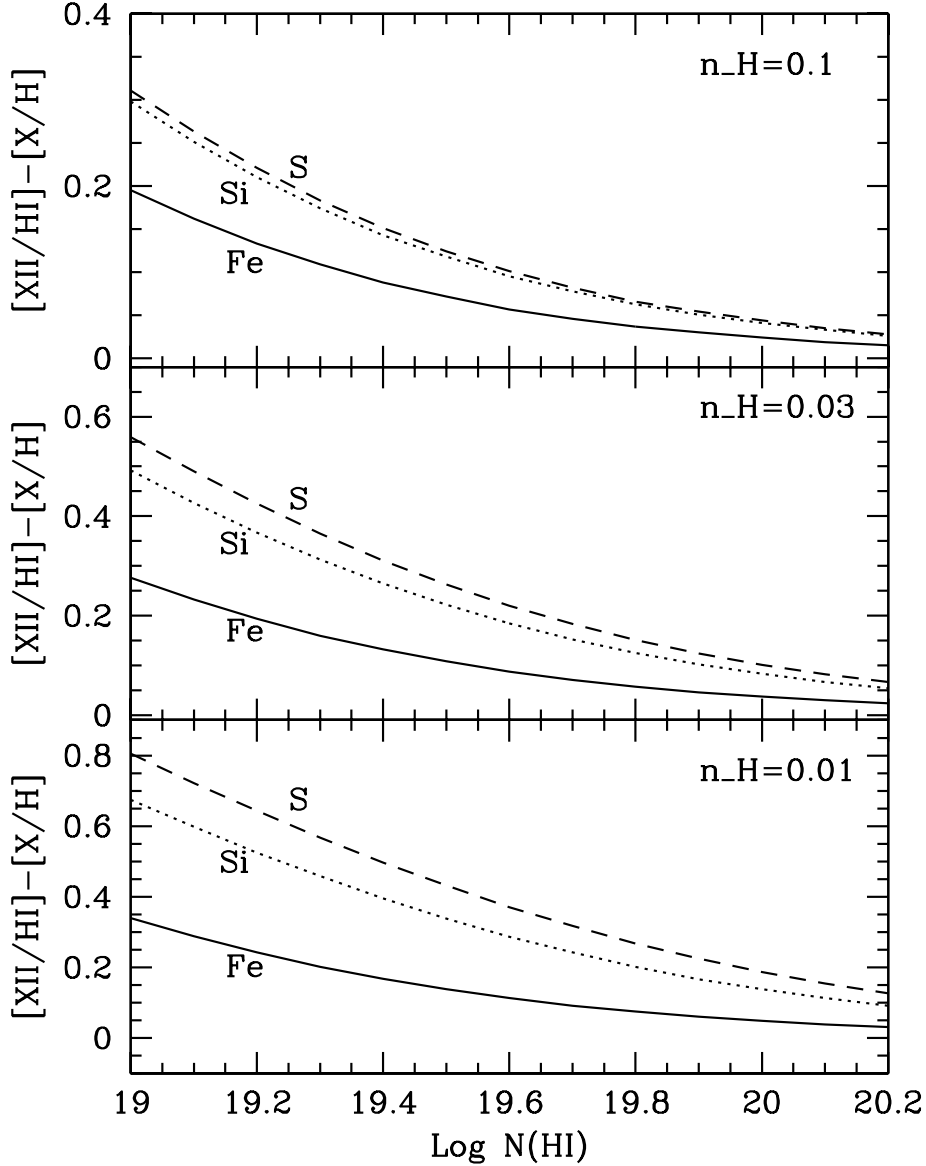}
\caption{CLOUDY models of the ionization correction for S, Si, and Fe 
versus \ion{H}{1}\ column density for three volume densities 
($n_{\rm H}=0.1$, 0.03, and 0.01 cm$^{-3}$).  The models feature an 
ionizing flux (photons cm$^{-2}$ s$^{-1}$) of log $\phi=5.5$ and a stellar
spectrum with
$T_{eff}=35,000$ K.  Other model details can be found in the text.
Plotted are the 
difference between the species' singly-ionized abundance relative to neutral
hydrogen, [\ion{X}{2}/\ion{H}{1}], and the true elemental abundance, [X/H].  We
find that the corrections necessary for \ion{O}{1}\ and 
\ion{N}{1}\ are negligible.}
\end{figure} 

\begin{figure}
\figurenum{17}
\plotone{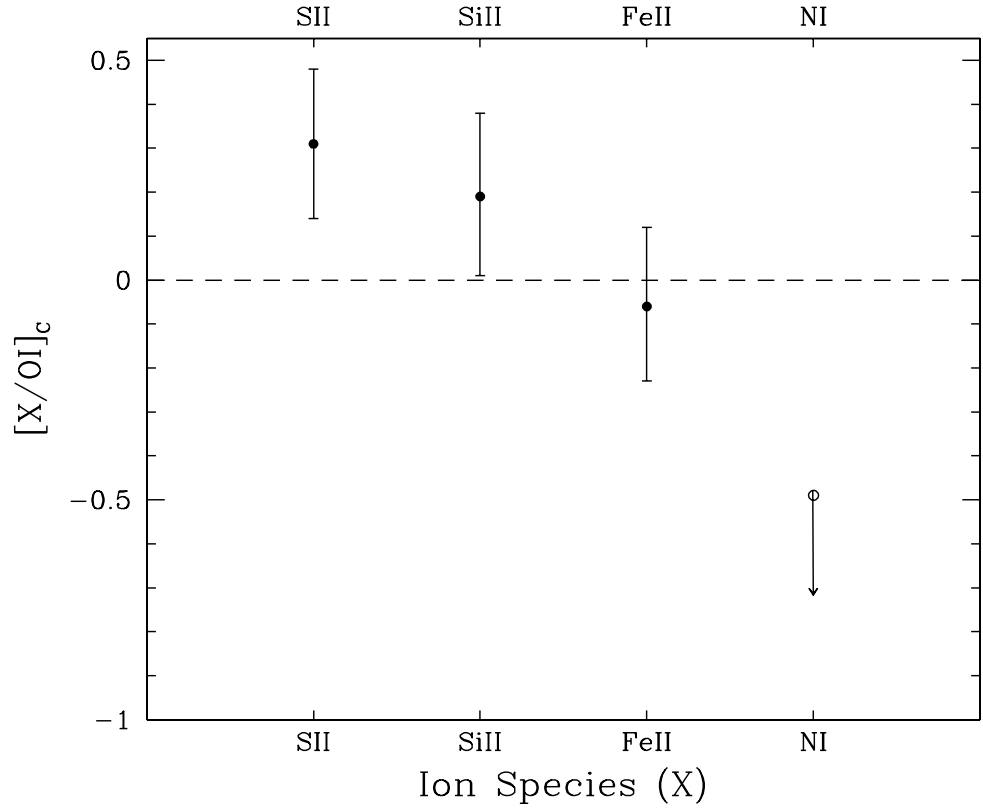}
\caption{Column density weighted mean abundances relative to \ion{O}{1}\ of 
\ion{S}{2}, 
\ion{Si}{2}, \ion{Fe}{2}, and \ion{N}{1} for the three sightlines through Complex C for which an \ion{O}{1}\ measurement could be made. 
The dashed line indicates a solar relative abundance.}
\end{figure} 

\clearpage
\begin{deluxetable}{lllcc}
\tablecolumns{5}
\tablewidth{0pc}
\tablecaption{SUMMARY OF {\it FUSE} OBSERVATIONS\tablenotemark{a} \label{t1}}
\tablehead{
\colhead{Sightline} & \colhead{Program ID} & \colhead{Observation Date(s)} & \colhead{$T_{exp}$(ks)} & \colhead{Number of Exposures} 
}
\startdata
Mrk 279     & P108 & 1999 Dec; 2000 Jan      & 91.9  & 27  \\
Mrk 290     & P107 & 2000 Mar                & 12.8  & 4   \\
Mrk 501     & P107 & 2000 Feb                & 11.6  & 4   \\
Mrk 817     & P108 & 2000 Feb, Dec; 2001 Feb & 190.0 & 63  \\
Mrk 876     & P107 & 1999 Oct                & 52.8  & 10  \\
PG 1259+593 & P108 & 2000 Feb; 2001 Jan, Mar & 610.6 & 215 \\
PG 1351+640 & P107, S601 & 2000 Jan; 2002 Feb & 119.0 & 34  \\
PG 1626+554 & P107 & 2000 Feb                & 8.3   & 2   \\
\enddata
\tablenotetext{a}{The {\it FUSE} spectrum covers the wavelength range of 905-1187 \AA.}
\end{deluxetable}

\clearpage
\begin{deluxetable}{lccccccc}
\tablecolumns{8}
\tablewidth{0pc}
\tablecaption{SUMMARY OF {\it HST} OBSERVATIONS \label{t2}}
\tablehead{
\colhead{} & \colhead{Proposal} & \colhead{Observation} & \colhead{} & \colhead{} & \colhead{Number of} & \colhead{} & \colhead{} \\
\colhead{Sightline} & \colhead{ID} & \colhead{Date} & \colhead{Instrument} & \colhead{$T_{exp}$(ks)} & \colhead{Exposures} & \colhead{$\lambda_{i}$(\AA)} & \colhead{$\lambda_{f}$(\AA)}
}
\startdata
Mrk 279     & 6593 & 1997 Jan & GHRS & 19.9 & 4  & 1222.6 & 1258.7 \\
Mrk 290     & 6590 & 1997 Jan & GHRS & 7.1  & 3  & 1232.9 & 1267.9 \\
Mrk 501     & 3584 & 1993 Feb & GHRS & 30.0 & 1  & 1222.5 & 1257.6 \\
Mrk 817     & 6593 & 1997 Jan & GHRS & 26.8 & 3  & 1222.7 & 1258.7 \\
Mrk 876     & 7295 & 1998 Sep & STIS & 2.3  & 1  & 1194.8 & 1249.4 \\
PG 1259+593 & 8695 & 2001 Jan & STIS & 81.3 & 34 & 1140.1 & 1729.6 \\
PG 1351+640 & 7345 & 2000 Aug & STIS & 14.7 & 2  & 1194.4 & 1299.0 \\
\enddata
\end{deluxetable}

\clearpage
\begin{deluxetable}{lcccccc}
\tablecolumns{9}
\tablewidth{0pc}
\tablecaption{LIMITS ON H$_{2}$ COLUMN DENSITIES AND FRACTIONS \label{t3}}
\tablehead{
\colhead{} & \multicolumn{2}{c}{$J=0$} & \multicolumn{2}{c}{$J=1$} & \colhead{} & \colhead{} \\
\colhead{Sightline} & \colhead{line} & \colhead{$W_{\lambda}$(m\AA)} & \colhead{line} & \colhead{$W_{\lambda}$(m\AA)} & \colhead{log $N$(H$_{2}$)\tablenotemark{a}} & \colhead{$f$(H$_{2}$)\tablenotemark{b}} }
\startdata
Mrk 279     & W(1-0) R(0) & $<36$ & W(1-0) Q(1) & $<36$ & $<14.24$ & $<1.0\times10^{-5}$ \\
Mrk 290     & W(0-0) R(0) & $<172$ & W(0-0) Q(1) & $<167$ & $<15.07$ & $<2.0\times10^{-5}$ \\
Mrk 501     & W(0-0) R(0) & $<118$ & W(0-0) Q(1) & $<121$ & $<14.93$ & $<1.1\times10^{-4}$ \\
Mrk 817     & W(0-0) R(0) & $<18$ & W(0-0) Q(1) & $<17$ & $<14.09$ & $<8.1\times10^{-6}$ \\
Mrk 876     & W(0-0) R(0) & $<51$ & W(0-0) Q(1) & $<51$ & $<14.56$ & $<3.1\times10^{-5}$ \\
PG 1259+593 & W(1-0) R(0) & $<38$ & W(0-0) Q(1) & $<25$ & $<14.25$ & $<4.2\times10^{-6}$ \\
PG 1351+640 & W(0-0) R(0) & $<83$ & W(0-0) Q(1) & $<83$ & $<14.77$ & $<1.9\times10^{-5}$ \\
PG 1626+554 & W(0-0) R(0) & $<166$ & W(0-0) Q(1) & $<160$ & $<15.06$ & $<3.7\times10^{-5}$ 
\enddata
\tablenotetext{a}{Column densities (in cm$^{-2}$) are calculated from the
indicated Werner lines 
assuming the optically thin case. All upper limits are 3$\sigma$.  We use the 
wavelengths and oscillator strengths of H$_{2}$ Werner lines from Abgrall et
al. (1993)}
\tablenotetext{b}{Molecular fraction is defined as
$f$(H$_{2}$)=2$N$(H$_{2}$)/[2$N$(H$_{2}$)+$N$(\ion{H}{1})].}
\end{deluxetable}

\clearpage
\begin{deluxetable}{lrlcccc}
\tablecolumns{7}
\tablewidth{0pc}
\tablecaption{SUMMARY OF MEASUREMENTS--MRK 279 SIGHTLINE\label{t4}}
\tablehead{
\colhead{} & \colhead{$\lambda$\tablenotemark{a}} & \colhead{} & \colhead{$W_{\lambda}$\tablenotemark{b}} & \colhead{log $N(X)$\tablenotemark{b}} & \colhead{} & \colhead{} \\
\colhead{Species} & \colhead{(\AA)} & \colhead{$f$\tablenotemark{a}} & \colhead{(m\AA)} & \colhead{($N$ in cm$^{-2}$)} & \colhead{log $A(X)_{\sun}$\tablenotemark{c}} & \colhead{[$X$/\ion{H}{1}]$_{\rm C}$}}
\startdata
\ion{H}{1}  & ...     \ \  & \ \ ...                     & ...       & $19.52^{+0.06}_{-0.06}$ & ...             & ...                            \\
            &              &                             &         & ($19.24^{+0.07}_{-0.08})$ &                 &               \\
\ion{N}{1}  & 1134.165 \ \ & \ \ 0.0152                  & $<26$     & $<14.24$                & $-4.08\pm0.06$  & $<$$-1.20$                     \\
\ion{O}{1}  & 924.950 \ \   & \ \ 0.00154                & $33\pm12$ & $15.64^{+0.35}_{-0.24}$ & $-3.17\pm0.06$  & $-0.71^{+0.36}_{-0.25}$         \\
            &               &                            &         & ($15.29^{+0.37}_{-0.23})$ &                & ($-0.78^{+0.38}_{-0.25}$)       \\
...         & 929.517 \ \   & \ \ 0.00229                & $44\pm12$ & ...                     & ...            & ...                             \\
...         & 936.630 \ \   & \ \ 0.00365                & $54\pm11$ & ...                     & ...            & ...                             \\
...         & 948.686 \ \   & \ \ 0.00632                & $92\pm10$ & ...                     & ...            & ...                             \\
...         & 1039.230 \ \  & \ \ 0.00919                & $101\pm7$ & ...                     & ...            & ...                             \\
\ion{Si}{2} & 1020.699 \ \  & \ \ 0.0164                 & $48\pm8$  & $14.67^{+0.11}_{-0.11}$ & $-4.44\pm0.01$  & $-0.41^{+0.13}_{-0.13}$        \\
            &               &                            &          & ($14.30^{+0.12}_{-0.13}$) &                & ($-0.50^{+0.14}_{-0.15}$)      \\
\ion{P}{2}  & 963.801  \ \  & \ \ 1.459                  & $<32$     & $<12.53$                & $-6.44\pm0.06$  & $<$$-0.55$                     \\
...         & 1152.818 \ \  & \ \ 0.245                  & $<24$      & ...                     & ...             & ...                           \\
\ion{S}{2}  & 1250.584 \ \  & \ \ 0.00543                & $19\pm6$  & $14.45^{+0.13}_{-0.18}$ & $-4.80\pm0.06$  & $-0.27^{+0.16}_{-0.20}$        \\
            &               &                            &          & ($14.21^{+0.15}_{-0.21}$) &                & ($-0.23^{+0.18}_{-0.23}$)      \\
\ion{S}{3}  & 1012.495 \ \  & \ \ 0.0442                 & $<25$     & $<13.79$ & $-4.80\pm0.06$ & $<$$-0.93$                    \\
\ion{Ar}{1} & 1048.220 \ \  & \ \ 0.263                  & $<21$     & $<12.97$                & $-5.60\pm0.06$  & $<$$-0.95$                     \\
\ion{Fe}{2} & 1096.877 \ \  & \ \ 0.032\tablenotemark{d} & $30\pm7$  & $14.16^{+0.21}_{-0.11}$ & $-4.50\pm0.01$  & $-0.86^{+0.22}_{-0.13}$        \\
            &               &                            &          & ($14.01^{+0.22}_{-0.13}$) &                & ($-0.73^{+0.23}_{-0.15}$)      \\
...         & 1121.975 \ \  & \ \ 0.0202\tablenotemark{d} & $33\pm7$ & ...                     & ...             & ...                            \\
...         & 1125.448 \ \  & \ \ 0.016\tablenotemark{d} & $29\pm8$  & ...                     & ...             & ...                            \\
...         & 1144.938 \ \  & \ \ 0.106\tablenotemark{d} & $78\pm8$  & ...                     & ...             & ...                            \\
\ion{Fe}{3} & 1122.524 \ \  & \ \ 0.0539                 & $33\pm9$  & $13.74^{+0.10}_{-0.14}$ & $-4.50\pm0.01$  & $-1.28^{+0.12}_{-0.15}$ 
\enddata
\tablenotetext{a}{Wavelengths and oscillator strengths are from Morton (2002) unless otherwise indicated.}
\tablenotetext{b}{Equivalent widths for Complex C are integrated over the velocity range -180 to -90 km s$^{-1}$.  All upper limits are 3$\sigma$. 
Column densities are calculated from a curve
of growth with doppler 
parameter, $b=9.7^{+5.0}_{-2.8}$ km s$^{-1}$, except for the cases of
\ion{S}{3}\ and \ion{Fe}{3} where we have assumed the lines to be 
optically thin.  Values in parenthesis are for the higher-velocity component 
in the integration range -180 to -120 km s$^{-1}$.} 
\tablenotetext{c}{Solar (meteoritic) abundance of element X 
from GS98.}
\tablenotetext{d}{Oscillator strength from Howk et al. (2000).}
\end{deluxetable}

\clearpage
\begin{deluxetable}{lrlcccc}
\tablecolumns{7}
\tablewidth{0pc}
\tablecaption{SUMMARY OF MEASUREMENTS--MRK 290 SIGHTLINE\label{t5}}
\tablehead{
\colhead{} & \colhead{$\lambda$\tablenotemark{a}} & \colhead{} & \colhead{$W_{\lambda}$\tablenotemark{b}} & \colhead{log $N(X)$} & \colhead{} & \colhead{} \\
\colhead{Species} & \colhead{(\AA)} & \colhead{$f$\tablenotemark{a}} & \colhead{(m\AA)} & \colhead{($N$ in cm$^{-2}$)} & \colhead{log $A(X)_{\sun}$\tablenotemark{c}} & \colhead{[$X$/\ion{H}{1}]$_{\rm C}$}}
\startdata
\ion{H}{1}\  & ...      \ \ & \ \ ...                     & ...         &  $20.09^{+0.02}_{-0.02}$    & ...               & ...                     \\
\ion{N}{1}\  & 1134.165 \ \ & \ \ 0.0152                  & $<183$      &  $<15.02$\tablenotemark{e}  & $-4.08\pm0.06$    & $<$$-0.99$              \\
\ion{O}{1}\  & 1039.230 \ \ & \ \ 0.00919             & $<151$          &  $<16.69$\tablenotemark{f}  & $-3.17\pm0.06$    & $<$$-0.23$              \\
\ion{Si}{2}\ & 1020.699 \ \ & \ \ 0.0164                  & $<141$      &  $<14.97$\tablenotemark{e}  & $-4.44\pm0.01$    & $<$$-0.68$              \\
\ion{S}{2}\ & 1253.801 \ \ & \ \ 0.0109                & $26\pm7$    &  $14.29^{+0.15}_{-0.14}$\tablenotemark{f} & $-4.80\pm0.06$  & $-1.00^{+0.16}_{-0.15}$ \\
...          & 1259.519 \ \ & \ \ 0.0166              & $36\pm12$   & ...  & ...  & ... \\
\ion{Ar}{1}\ & 1048.220 \ \ & \ \ 0.263                   & $<133$      &  $<13.72$\tablenotemark{e}           & $-5.60\pm0.06$    & $<$$-0.77$     \\
\ion{Fe}{2}\ & 1144.938 \ \ & \ \ 0.106\tablenotemark{d}  & $<161$      &  $<15.46$\tablenotemark{f}           & $-4.50\pm0.01$    & $<$$-0.13$ \\
\enddata
\tablenotetext{a}{Wavelengths and oscillator strength are from Morton (2002) unless otherwise indicated.}
\tablenotetext{b}{Equivalent widths for Complex C are integrated over the velocity range -165 to -80 km s$^{-1}$.  All upper limits are 3$\sigma$ except for \ion{O}{1} $\lambda$1039.230, where the limit is taken as the measured value since the line is blended with Galactic H$_{2}$ absorption.} 
\tablenotetext{c}{Solar (meteoritic) abundance of element X 
from GS98.}
\tablenotetext{d}{Oscillator strength from Howk et al. (2000).}
\tablenotetext{e}{Assuming the optically thin case.}
\tablenotetext{f}{Assuming the equivalent width is related to the column density through a curve of growth with doppler parameter, $b=10$ km s$^{-1}$.}
\end{deluxetable}

\clearpage
\begin{deluxetable}{lrlcccc}
\tablecolumns{7}
\tablewidth{0pc}
\tablecaption{SUMMARY OF MEASUREMENTS--MRK 501 SIGHTLINE\label{t6}}
\tablehead{
\colhead{} & \colhead{$\lambda$\tablenotemark{a}} & \colhead{} & \colhead{$W_{\lambda}$\tablenotemark{b}} & \colhead{log $N(X)$} & \colhead{} & \colhead{} \\
\colhead{Species} & \colhead{(\AA)} & \colhead{$f$\tablenotemark{a}} & \colhead{(m\AA)} & \colhead{($N$ in cm$^{-2}$)} & \colhead{log $A(X)_{\sun}$\tablenotemark{c}} & \colhead{[$X$/\ion{H}{1}]$_{\rm C}$}}
\startdata
\ion{H}{1}\  & ...      \ \ & \ \ ...                     & ...         &  $19.19^{+0.03}_{-0.03}$ & ...                     & ...                     \\
\ion{N}{1}\  & 1134.165 \ \ & \ \ 0.0152                  & $<101$      &  $<14.77$\tablenotemark{e}           & $-4.08\pm0.06$        & $<$$-0.34$    \\
\ion{O}{1}\  & 1039.230 \ \ & \ \ 0.00919                 & $<94$       &  $<15.48$\tablenotemark{f}           & $-3.17\pm0.06$        & $<$$-0.54$    \\
\ion{Si}{2}\ & 1020.699 \ \ & \ \ 0.0164                  & $<90$       &  $<14.78$\tablenotemark{e}           & $-4.44\pm0.01$        & $<0.03$       \\
\ion{S}{2}\  & 1253.801 \ \ & \ \ 0.0109                  & $<42$       &  $<14.44$\tablenotemark{e}           & $-4.80\pm0.06$        & $<0.05$       \\
\ion{Ar}{1}\ & 1048.220 \ \ & \ \ 0.263                   & $<91$       &  $<13.55$\tablenotemark{e}           & $-5.60\pm0.06$        & $<$$-0.04$    \\
\ion{Fe}{2}\ & 1144.938 \ \ & \ \ 0.106\tablenotemark{d}  & $<92$       &  $<14.24$\tablenotemark{f}           & $-4.50\pm0.01$        & $<$$-0.45$    \\
\enddata
\tablenotetext{a}{Wavelengths and oscillator strength are from Morton (2002) unless otherwise indicated.}
\tablenotetext{b}{Equivalent widths for Complex C are integrated over the velocity range -135 to -65 km s$^{-1}$.  All upper limits are 3$\sigma$.} 
\tablenotetext{c}{Solar (meteoritic) abundance of element X 
from GS98.}
\tablenotetext{d}{Oscillator strength from Howk et al. (2000).}
\tablenotetext{e}{Assuming the optically thin case.}
\tablenotetext{f}{Assuming the equivalent width is related to the column density through a curve of growth with doppler parameter, $b=10$ km s$^{-1}$.}
\end{deluxetable}

\clearpage
\begin{deluxetable}{lrlcccc}
\tablecolumns{7}
\tablewidth{0pc}
\tablecaption{SUMMARY OF MEASUREMENTS--MRK 817 SIGHTLINE\label{t7}}
\tablehead{
\colhead{} & \colhead{$\lambda$\tablenotemark{a}} & \colhead{} & \colhead{$W_{\lambda}$\tablenotemark{b}} & \colhead{log $N(X)$\tablenotemark{c}} & \colhead{} & \colhead{} \\
\colhead{Species} & \colhead{(\AA)} & \colhead{$f$\tablenotemark{a}} & \colhead{(m\AA)} & \colhead{($N$ in cm$^{-2}$)} & \colhead{log $A(X)_{\sun}$\tablenotemark{d}} & \colhead{[$X$/\ion{H}{1}]$_{\rm C}$}}
\startdata
\ion{H}{1}  & ...     \ \  & \ \ ...                     & ...       & $19.48^{+0.01}_{-0.01}$ & ...            & ...                            \\
\ion{N}{1}  & 1134.165 \ \ & \ \ 0.0152                  & $<18$     & $<14.05$                  & $-4.08\pm0.06$ & $<$$-1.35$                     \\
\ion{O}{1}  & 924.950  \ \ & \ \ 0.00154                 & $49\pm9$  & $15.72^{+0.24}_{-0.16}$   & $-3.17\pm0.06$ & $-0.59^{+0.25}_{-0.17}$        \\
...         & 929.517  \ \ & \ \ 0.00229                 & $50\pm9$  & ...                       & ...            & ...                            \\
...         & 936.630  \ \ & \ \  0.00365                & $73\pm9$  & ...                       & ...            & ...                            \\
...         & 948.686  \ \ & \ \  0.00632                & $92\pm8$  & ...                       & ...            & ...                            \\
...         & 1039.230 \ \ & \ \  0.00919                & $115\pm5$ & ...                       & ...            & ...                            \\
\ion{Si}{2} & 1020.699 \ \ & \ \  0.0164                 & $36\pm5$  & $14.48^{+0.07}_{-0.08}$   & $-4.44\pm0.01$ & $-0.56^{+0.07}_{-0.08}$        \\
\ion{P}{2}  & 963.801  \ \ & \ \  1.459                  & $<55$     & $<12.84$                  & $-6.44\pm0.06$  & $<$$-0.20$                    \\
...         & 1152.818 \ \ & \ \  0.245                  & $<19$     & ...                       & ...             & ...                           \\
\ion{S}{2}  & 1250.584 \ \ & \ \  0.00543                & $16\pm3$  & $14.34^{+0.05}_{-0.05}$   & $-4.80\pm0.06$  & $-0.34^{+0.08}_{-0.08}$       \\
...         & 1253.801 \ \ & \ \  0.0109                 & $28\pm3$  & ...                       & ...             & ...                           \\
\ion{S}{3}  & 1012.495 \ \ & \ \  0.0442                 & $<24$     & $<13.78$ & $-4.80\pm0.06$  & $<$$-0.90$                    \\
\ion{Ar}{1} & 1048.220 \ \ & \ \  0.263                  & $<15$     & $<12.81$                  & $-5.60\pm0.06$  & $<$$-1.07$                    \\
\ion{Fe}{2} & 1063.176 \ \ & \ \  0.0548                 & $66\pm4$  & $14.31^{+0.11}_{-0.08}$   & $-4.50\pm0.01$  & $-0.67^{+0.11}_{-0.08}$       \\
...         & 1096.877 \ \ & \ \  0.032\tablenotemark{e} & $46\pm5$  & ...                       & ...             & ...                           \\
...         & 1121.975 \ \ & \ \  0.0202\tablenotemark{e} & $41\pm5$ & ...                       & ...             & ...                           \\
...         & 1125.448 \ \ & \ \  0.016\tablenotemark{e} & $32\pm6$  & ...                       & ...             & ...                           \\
...         & 1143.226 \ \ & \ \  0.0177\tablenotemark{e} & $34\pm6$ & ...                       & ...             & ...                           \\
...         & 1144.938 \ \ & \ \  0.106\tablenotemark{e} & $104\pm6$ & ...                       & ...             & ...                           \\
\ion{Fe}{3} & 1122.524 \ \ & \ \  0.0539                 & $30\pm6$  & $13.70^{+0.08}_{-0.10}$ & $-4.50\pm0.01$  & $-1.28^{+0.08}_{-0.10}$    
\enddata
\tablenotetext{a}{Wavelengths and oscillator strengths are from Morton (2002) unless otherwise indicated.}
\tablenotetext{b}{Equivalent widths for Complex C are integrated over the velocity range -140 to -80 km s$^{-1}$.  All upper limits are 3$\sigma$ levels except for the \ion{S}{3} $\lambda$1012.5 and \ion{P}{2} $\lambda963.8$ lines, where the upper limit is taken as the measured value since the Complex C component is blended with other absorption features.} 
\tablenotetext{c}{All column densities are calculated from a curve
of growth with doppler 
parameter, $b=10.8^{+3.2}_{-2.2}$ km s$^{-1}$, except for the cases of
\ion{S}{3}\ and \ion{Fe}{3} where we have assumed the lines to be 
optically thin.} 
\tablenotetext{d}{Solar (meteoritic) abundance of element X 
from GS98.}
\tablenotetext{e}{Oscillator strength from Howk et al. (2000).}
\end{deluxetable}

\clearpage
\begin{deluxetable}{lrlcccc}
\tablecolumns{7}
\tablewidth{0pc}
\tablecaption{SUMMARY OF MEASUREMENTS--MRK 876 SIGHTLINE\label{t8}}
\tablehead{
\colhead{} & \colhead{$\lambda$\tablenotemark{a}} & \colhead{} & \colhead{$W_{\lambda}$\tablenotemark{b}} & \colhead{log $N(X)$\tablenotemark{b}} & \colhead{} & \colhead{} \\
\colhead{Species} & \colhead{(\AA)} & \colhead{$f$\tablenotemark{a}} & \colhead{(m\AA)} & \colhead{($N$ in cm$^{-2}$)} & \colhead{log $A(X)_{\sun}$\tablenotemark{c}} & \colhead{[$X$/\ion{H}{1}]$_{\rm C}$}}
\startdata
\ion{H}{1}  & ...      \ \ & \ \ ...                      & ...        & $19.37^{+0.02}_{-0.02}$ & ...                 & ...                   \\
            &              &                              &            & ($19.26^{+0.03}_{-0.03}$) &                    &                       \\
\ion{N}{1}  & 1199.550 \ \ & \ \ 0.130                    & $136\pm21$ & $14.20^{+0.15}_{-0.14}$ & $-4.08\pm0.06$  & $-1.09^{+0.16}_{-0.15}$   \\
            &              &                              &            & ($13.99^{+0.12}_{-0.10}$) &               & ($-1.19^{+0.14}_{-0.12}$)  \\                        
...         & 1134.165 \ \ & \ \ 0.0152                   & $<49$      & ...                     & ...             & ...                       \\
\ion{O}{1}  &  936.630 \ \ & \ \ 0.00365                  & $<108$     & $<15.88$                & $-3.17\pm0.06$  & $<$$-0.32$                  \\
            &              &                              &            & ($15.55^{+0.42}_{-0.28}$) &                    & ($-0.54^{+0.43}_{-0.29}$) \\
\ion{Si}{2} & 1020.699 \ \ & \ \  0.0164                  & $55\pm14$  & $14.67^{+0.13}_{-0.16}$ & $-4.44\pm0.01$  & $-0.26^{+0.13}_{-0.16}$   \\
            &              &                              &            & ($14.53^{+0.11}_{-0.13}$) &                    & ($-0.29^{+0.11}_{-0.13}$) \\
\ion{P}{2}  & 1152.818 \ \ & \ \  0.245                   & $<40$      & $<13.21$                & $-6.44\pm0.06$  & $<0.28$                   \\
\ion{Ar}{1} & 1048.220 \ \ & \ \  0.263                   & $<43$      & $<13.30$                & $-5.60\pm0.06$  & $<$$-0.47$                \\
\ion{Fe}{2} & 1121.975 \ \ & \ \  0.0202\tablenotemark{d} & $52\pm12$  & $14.45^{+0.15}_{-0.12}$ & $-4.50\pm0.01$  & $-0.42^{+0.15}_{-0.12}$   \\
            &              &                              &            & ($14.41^{+0.09}_{-0.08}$) &                    & ($-0.35^{+0.10}_{-0.09}$) \\
...         & 1125.448 \ \ & \ \  0.016\tablenotemark{d}  & $37\pm12$  & ...                     & ...             & ...                       \\
...         & 1143.226 \ \ & \ \  0.0177\tablenotemark{d} & $53\pm14$  & ...                     & ...             & ...                       \\
...         & 1144.938 \ \ & \ \  0.106\tablenotemark{d}  & $149\pm13$ & ...                     & ...             & ...                       \\
\ion{Fe}{3} & 1122.524 \ \ & \ \  0.0539                  & $<48$      & $<13.90$ & $-4.50\pm0.01$ & $<$$-0.97$    
\enddata
\tablenotetext{a}{Wavelengths and oscillator strengths are from Morton (2002) unless otherwise indicated.}
\tablenotetext{b}{Equivalent widths for Complex C are integrated over the velocity range -210 to -95 km s$^{-1}$. All upper limits are 3$\sigma$ levels. 
Column densities are calculated from a curve
of growth with doppler 
parameter, $b=16.1^{+7.4}_{-3.9}$ km s$^{-1}$, except for the case of
\ion{Fe}{3} where we have assumed the line to be 
optically thin.  Values in parenthesis are for the lower-velocity component in
the integration range -150 to -95 km s$^{-1}$.  For the higher-velocity 
component in the integration range -210 to -150 km s$^{-1}$, 
we measure [\ion{Fe}{2}/\ion{H}{1}]$_{\rm C}=-0.71^{+0.14}_{-0.15}$.}
\tablenotetext{c}{Solar (meteoritic) abundance of element X 
from GS98.}
\tablenotetext{d}{Oscillator strength from Howk et al. (2000).}
\end{deluxetable}

\clearpage
\begin{deluxetable}{lrlcccc}
\tablecolumns{7}
\tablewidth{0pc}
\tablecaption{SUMMARY OF MEASUREMENTS--PG 1259+593 SIGHTLINE\label{t9}}
\tablehead{
\colhead{} & \colhead{$\lambda$\tablenotemark{a}} & \colhead{} & \colhead{$W_{\lambda}$\tablenotemark{b}} & \colhead{log $N(X)$\tablenotemark{c}} & \colhead{} & \colhead{} \\
\colhead{Species} & \colhead{(\AA)} & \colhead{$f$\tablenotemark{a}} & \colhead{(m\AA)} & \colhead{($N$ in cm$^{-2}$)} & \colhead{log $A(X)_{\sun}$\tablenotemark{d}} & \colhead{[$X$/\ion{H}{1}]$_{\rm C}$}}
\startdata
\ion{H}{1}  & ...      \ \ & \ \ ...                     & ...        & $19.92^{+0.01}_{-0.01}$ & ...            & ...                            \\
\ion{N}{1}  & 1134.165 \ \ & \ \ 0.0152                  & $20\pm7$   & $14.02^{+0.19}_{-0.12}$ & $-4.08\pm0.06$ & $-1.82^{+0.20}_{-0.13}$        \\
...         & 1199.550 \ \ & \ \ 0.130                   & $79\pm13$  & ...                     & ...            & ...                            \\
\ion{O}{1}  & 924.950  \ \ & \ \ 0.00154                 & $55\pm11$  & $15.75^{+0.18}_{-0.24}$ & $-3.17\pm0.06$ & $-1.00^{+0.19}_{-0.25}$        \\
...         & 929.517  \ \ & \ \ 0.00229                 & $54\pm10$  & ...                     & ...            & ...                            \\
...         & 936.630  \ \ & \ \ 0.00365                 & $73\pm10$  & ...                     & ...            & ...                            \\
...         & 948.686  \ \ & \ \ 0.00632                 & $71\pm12$  & ...                     & ...            & ...                            \\
...         & 1039.230 \ \ & \ \ 0.00919                 & $97\pm6$   & ...                     & ...            & ...                            \\
...         & 1302.169 \ \ & \ \ 0.0519                  & $203\pm8$  & ...                     & ...            & ...                            \\
\ion{Al}{2} & 1670.787 \ \ & \ \ 1.828                   & $193\pm13$ & $13.45^{+0.17}_{-0.16}$ & $-5.51\pm0.01$  & $-0.96^{+0.17}_{-0.16}$       \\
\ion{Si}{2} & 1020.699 \ \ & \ \ 0.0164                  & $50\pm7$   & $14.67^{+0.20}_{-0.15}$ & $-4.44\pm0.01$  & $-0.81^{+0.20}_{-0.15}$       \\
...         & 1193.290 \ \ & \ \ 0.585                   & $174\pm13$ & ...                     & ...             & ...                           \\
...         & 1304.370 \ \ & \ \ 0.0917                  & $134\pm8$  & ...                     & ...             & ...                           \\
...         & 1526.707 \ \ & \ \ 0.132                   & $178\pm10$ & ...                     & ...             & ...                           \\
\ion{P}{2}  & 1152.818 \ \ & \ \ 0.245                   & $<23$      & $<12.96$                & $-6.44\pm0.06$  & $<$$-0.52$                    \\
\ion{S}{2}  & 1253.801 \ \ & \ \ 0.0109                  & $30\pm10$  & $14.38^{+0.12}_{-0.11}$ & $-4.80\pm0.06$  & $-0.74^{+0.13}_{-0.13}$       \\
...         & 1259.519 \ \ & \ \ 0.0166                  & $44\pm10$  & ...                     & ...             & ...                           \\
\ion{S}{3}  & 1012.495 \ \ & \ \ 0.0442                  & $<23$      & $<13.76$ & $-4.80\pm0.06$ & $<$$-1.36$                   \\
\ion{Ar}{1} & 1048.220 \ \ & \ \ 0.263                   & $27\pm7$   & $13.10^{+0.12}_{-0.15}$ & $-5.60\pm0.06$  & $-1.22^{+0.13}_{-0.16}$       \\
\ion{Fe}{2} & 1055.262 \ \ & \ \ 0.0075\tablenotemark{e} & $27\pm7$   & $14.40^{+0.17}_{-0.08}$ & $-4.50\pm0.01$  & $-1.02^{+0.17}_{-0.08}$       \\
...         & 1063.176 \ \ & \ \ 0.0548                  & $75\pm7$   & ...                     & ...             & ...                           \\
...         & 1096.877 \ \ & \ \ 0.032\tablenotemark{e}  & $49\pm6$   & ...                     & ...             & ...                           \\
...         & 1121.975 \ \ & \ \ 0.0202\tablenotemark{e} & $44\pm6$   & ...                     & ...             & ...                           \\
...         & 1125.448 \ \ & \ \ 0.016\tablenotemark{e}  & $41\pm6$   & ...                     & ...             & ...                           \\
...         & 1143.226 \ \ & \ \ 0.0177\tablenotemark{e} & $41\pm7$   & ...                     & ...             & ...                           \\
...         & 1144.938 \ \ & \ \ 0.106\tablenotemark{e}  & $88\pm7$   & ...                     & ...             & ...                           \\
...         & 1608.451 \ \ & \ \ 0.0580                  & $130\pm16$ & ...                     & ...             & ...                           \\
\ion{Fe}{3} & 1122.524 \ \ & \ \ 0.0539                  & $<26$      & $<13.64$ & $-4.50\pm0.01$ & $<$$-1.78$     
\enddata
\tablenotetext{a}{Wavelengths and oscillator strengths are from Morton (2002) unless otherwise indicated.}
\tablenotetext{b}{Equivalent widths for Complex C are integrated over the velocity range -155 to -95 km s$^{-1}$.  All upper limits are 3$\sigma$ levels.} 
\tablenotetext{c}{All column densities are calculated from a curve
of growth with doppler 
parameter, $b=10.0^{+1.9}_{-1.5}$ km s$^{-1}$, except for the cases of
\ion{S}{3} and \ion{Fe}{3} where we have assumed the lines to be 
optically thin.} 
\tablenotetext{d}{Solar (meteoritic) abundance of element X 
from GS98.}
\tablenotetext{e}{Oscillator strength from Howk et al. (2000).}
\end{deluxetable}

\clearpage
\begin{deluxetable}{lrlcccc}
\tablecolumns{7}
\tablewidth{0pc}
\tablecaption{SUMMARY OF MEASUREMENTS--PG 1351+640 SIGHTLINE\label{t10}}
\tablehead{
\colhead{} & \colhead{$\lambda$\tablenotemark{a}} & \colhead{} & \colhead{$W_{\lambda}$\tablenotemark{b}} & \colhead{log $N(X)$\tablenotemark{c}} & \colhead{} & \colhead{} \\
\colhead{Species} & \colhead{(\AA)} & \colhead{$f$\tablenotemark{a}} & \colhead{(m\AA)} & \colhead{($N$ in cm$^{-2}$)} & \colhead{log $A(X)_{\sun}$\tablenotemark{d}} & \colhead{[$X$/\ion{H}{1}]$_{\rm C}$}}
\startdata
\ion{H}{1}  & ...      \ \ & \ \ ...                     & ...        & $19.80^{+0.02}_{-0.02}$ & ...             & ...                          \\
\ion{N}{1}  & 1134.165 \ \ & \ \ 0.0152                  & $<83$      & $<14.86$                & $-4.08\pm0.06$  & $<$$-0.86$                   \\
\ion{O}{1}  & 1039.230 \ \ & \ \ 0.00919                 & $<178$     & $<16.30$                & $-3.17\pm0.06$  & $<$$-0.33$                   \\
\ion{Si}{2} & 1020.699 \ \ & \ \ 0.0164                  & $84\pm19$  & $14.97^{+0.17}_{-0.18}$ & $-4.44\pm0.01$  & $-0.39^{+0.17}_{-0.18}$      \\
\ion{S}{2}  & 1253.801 \ \ & \ \ 0.0109                  & $<42$      & $<14.51$                & $-4.80\pm0.06$  & $<$$-0.49$                   \\
\ion{Ar}{1} & 1048.220 \ \ & \ \ 0.263                   & $<59$      & $<13.50$                & $-5.60\pm0.06$  & $<$$-0.70$                   \\
\ion{Fe}{2} & 1143.226 \ \ & \ \ 0.0177\tablenotemark{e} & $89\pm19$ & $14.85^{+0.26}_{-0.17}$ & $-4.50\pm0.01$  & $-0.45^{+0.26}_{-0.17}$      \\
...         & 1144.938 \ \ & \ \ 0.106\tablenotemark{e}  & $171\pm19$ & ...                     & ...             & ...                          
\enddata
\tablenotetext{a}{Wavelengths and oscillator strengths are from Morton (2002) unless otherwise indicated.}
\tablenotetext{b}{Equivalent widths for Complex C are integrated over the velocity range -190 to -95 km s$^{-1}$.  All upper limits are 3$\sigma$ levels except for \ion{O}{1} $\lambda$1039.230, where the limit is taken as the measured value since the line is blended with Galactic H$_{2}$ absorption.} 
\tablenotetext{c}{All column densities are calculated from a curve
of growth with doppler 
parameter, $b=13.7^{+3.3}_{-2.5}$ km s$^{-1}$.} 
\tablenotetext{d}{Solar (meteoritic) abundance of element X 
from GS98.}
\tablenotetext{e}{Oscillator strength from Howk et al. (2000).}
\end{deluxetable}

\begin{deluxetable}{lcccccc}
\tablecolumns{7}
\tablewidth{0pc}
\tablecaption{SUMMARY OF MEASUREMENTS--PG 1626+554\label{t11}}
\tablehead{
\colhead{} & \colhead{$\lambda$\tablenotemark{a}} & \colhead{} & \colhead{$W_{\lambda}$\tablenotemark{b}} & \colhead{log $N(X)$} & \colhead{} & \colhead{} \\
\colhead{Species} & \colhead{(\AA)} & \colhead{$f$\tablenotemark{a}} & \colhead{(m\AA)} & \colhead{($N$ in cm$^{-2}$)} & \colhead{log $A(X)_{\sun}$\tablenotemark{c}} & \colhead{[$X$/\ion{H}{1}]$_{\rm C}$}}
\startdata
\ion{H}{1}\  & ...      \ \ & \ \ ...                    & ...       & $19.79^{+0.03}_{-0.03}$   & ...                     & ...                     \\
\ion{N}{1}\  & 1134.165 \ \ & \ \ 0.0152                 & $<118$    & $<14.83$\tablenotemark{e} & $-4.08\pm0.06$ &  $<$$-0.88$              \\
\ion{O}{1}\  & 1039.230 \ \ & \ \ 0.00919                & $<114$    & $<15.83$\tablenotemark{f} & $-3.17\pm0.06$ &  $<$$-0.79$              \\
\ion{Si}{2}\ & 1020.699 \ \ & \ \ 0.0164                 & $<135$    & $<14.95$\tablenotemark{e} & $-4.44\pm0.01$ &  $<$$-0.40$              \\
\ion{Ar}{1}\ & 1048.220 \ \ & \ \ 0.263                  & $<128$    & $<13.70$\tablenotemark{e} & $-5.60\pm0.06$ &  $<$$-0.49$              \\
\ion{Fe}{2}\ & 1144.938 \ \ & \ \ 0.106\tablenotemark{d} & $<116$    & $<14.57$\tablenotemark{f} & $-4.50\pm0.01$ &  $<$$-0.72$
\enddata
\tablenotetext{a}{Wavelengths and oscillator strengths are from Morton (2002) unless otherwise indicated.}
\tablenotetext{b}{Equivalent widths for Complex C are integrated over the velocity range -155 to -75 km s$^{-1}$.  All upper limits are 3$\sigma$ except for \ion{O}{1} $\lambda$1039.230, where the limit is taken as the measured value since the line is blended with Galactic H$_{2}$ absorption.} 
\tablenotetext{c}{Solar (meteoritic) abundance of element X 
from GS98.}
\tablenotetext{d}{Oscillator strength from Howk et al. (2000).}
\tablenotetext{e}{Assuming the optically thin case.}
\tablenotetext{f}{Assuming the equivalent width is related to the column density through a curve of growth with doppler parameter, $b=10$ km s$^{-1}$.}
\end{deluxetable}

\clearpage
\begin{deluxetable}{lcccccc}
\tablecolumns{7}
\tablewidth{0pc}
\tablecaption{COMPLEX C ION ABUNDANCES\label{t12}}
\tablehead{\colhead{Sightline} & \colhead{log $N_{HI}$(cm$^{-2}$)} & \colhead{[\ion{O}{1}/\ion{H}{1}]$_{\rm C}$} & \colhead{[\ion{S}{2}/\ion{H}{1}]$_{\rm C}$} & \colhead{[\ion{Fe}{2}/\ion{H}{1}]$_{\rm C}$} & \colhead{[\ion{Si}{2}/\ion{H}{1}]$_{\rm C}$} & \colhead{[\ion{N}{1}/\ion{H}{1}]$_{\rm C}$}}
\startdata
Mrk 279     & $19.52^{+0.06}_{-0.06}$ & $-0.71^{+0.36}_{-0.25}$ & $-0.27^{+0.16}_{-0.20}$ & $-0.86^{+0.22}_{-0.13}$ & $-0.41^{+0.13}_{-0.13}$ & $<$$-1.20$ \\
Mrk 290     & $20.09^{+0.02}_{-0.02}$ & $<$$-0.23$              & $-1.00^{+0.16}_{-0.15}$ & $<$$-0.13$              & $<$$-0.68$              & $<$$-0.99$ \\
Mrk 501     & $19.19^{+0.03}_{-0.03}$ & $<$$-0.54$              & $<0.05$               & $<$$-0.45$              & $<0.03$                & $<$$-0.34$ \\
Mrk 817     & $19.48^{+0.01}_{-0.01}$ & $-0.59^{+0.25}_{-0.17}$ & $-0.34^{+0.08}_{-0.08}$ & $-0.67^{+0.11}_{-0.08}$ & $-0.56^{+0.07}_{-0.08}$  & $<$$-1.35$ \\
Mrk 876     & $19.37^{+0.02}_{-0.02}$ & $<$$-0.32$            & ...                     & $-0.42^{+0.15}_{-0.12}$ & $-0.26^{+0.13}_{-0.16}$  & $-1.09^{+0.16}_{-0.15}$ \\
PG 1259+593 & $19.92^{+0.01}_{-0.01}$ & $-1.00^{+0.19}_{-0.25}$ & $-0.74^{+0.13}_{-0.13}$ & $-1.02^{+0.17}_{-0.08}$ & $-0.81^{+0.20}_{-0.15}$  & $-1.82^{+0.20}_{-0.13}$ \\
PG 1351+640 & $19.80^{+0.02}_{-0.02}$ & $<$$-0.33$              & $<$$-0.49$              & $-0.45^{+0.26}_{-0.17}$ & $-0.39^{+0.17}_{-0.18}$  & $<$$-0.86$ \\
PG 1626+554 & $19.79^{+0.03}_{-0.03}$ & $<$$-0.79$              & ...                     & $<$$-0.72$              & $<$$-0.40$               & $<$$-0.88$ 
\enddata
\end{deluxetable}

\end{document}